\documentclass{article}
\usepackage[utf8]{inputenc}

\usepackage[margin=1.in]{geometry}

\usepackage{rotating}

\usepackage{comment}
\usepackage{subcaption}
\usepackage{tikz}
\usepackage{pgfplots}
\usepackage{amsmath,amssymb}
\usepackage{hyperref}
\usepackage{caption}
\usepackage{authblk}
\usepackage{multirow}
\usepackage{soul}

\usepackage{fancyhdr}
\newcommand{\D}[1]{\ensuremath{\operatorname{d}\!{#1}}} 

\title{Forming the Trappist-1 system in two steps during the recession of the disc inner edge}
\author[1]{Gabriele Pichierri}
\author[2,3]{Alessandro Morbidelli}
\author[1]{Konstantin Batygin}
\author[4,5]{Ramon Brasser}
\affil[1]{\small Division of Geological and Planetary Sciences California Institute of Technology, Pasadena, CA 91125, USA (corresponding author: G.P.: gabe@caltech.edu)}
\affil[2]{Laboratoire Lagrange, Université Cote d’Azur, CNRS, Observatoire de la Cote d’Azur, Nice, France}
\affil[3]{Collège de France, CNRS, PSL Univ., Sorbonne Univ., Paris, 75014, France}
\affil[4]{Origins Research Institute, Research Centre for Astronomy and Earth Sciences, Konkoly Thege Miklos St. 15-17, H-1121 Budapest}
\affil[5]{Centre for Planetary Habitability, University of Oslo; Sem Saelands vei 2A, N-0315 Oslo, Norway}
\date{}

\begin{document}

\maketitle

Trappist-1 hosts 7 planets where period ratios of neighbouring pairs are close to the 8:5, 5:3, 3:2, 3:2, 4:3, and 3:2 ratios in increasing distance from the star. The Laplace angles associated with neighbouring triplets are observed to be librating, proving the resonant nature of the system. This compact, resonant configuration is a manifest sign of disc-driven migration; however, the preferred outcome of such evolution is the establishment of first-order resonances, not the high-order resonances observed in the inner system. 
Here, we explain the observed orbital configuration in a model that is largely independent on the specific disc migration and orbital circularisation efficiencies. 
Together with migration, the two key elements of our model are: i) the inner border of the protoplanetary disc receded with time; and ii) the system was initially separated in two sub-systems. Specifically, the inner b, c d and e planets were initially placed in a 3:2 resonance chain and then evolved to the 8:5 -- 5:3 commensurability between planets b, c and d under the effect of the recession of the inner edge of the disc, whereas the outer planets migrated to the inner edge at a later time, establishing the remaining resonances. Our results pivot on the dynamical role of the presently unobservable recession of the inner edge of protoplanetary discs. They also reveal the role of recurring phases of convergent migration followed by resonant repulsion with associated orbital circularisation when resonant chains interact with migration barriers.\\

{\bf Introduction.} The main difficulty in explaining the observed \cite{2017NatAs...1E.129L,2021PSJ.....2....1A} orbital configuration of the Trappist-1 system consists in reproducing the inner planets' 8:5 and 5:3 period ratios combined with the outer planets' vicinity to first order 3:2 and 4:3 resonances. 
$N$-body simulations of the full 7-planet Trappist-1 system undergoing classical \cite{2012ARA&A..50..211K,2017ApJ...840L..19T} disc-driven type-I migration show that the natural outcome of this scenario is the formation of a simpler first-order 3:2 -- 3:2 -- 3:2 -- 3:2 -- 4:3 -- 3:2 resonant chain \cite{2022A&A...658A.170T,2022MNRAS.515.2373B}.  
\cite{2018MNRAS.476.5032P} argued for a two-subsystem structure based on the libration of Laplace angles, but noted that the observed period ratios were obtained via disc-driven migration by adjusting migration and eccentricity damping timescales.
\cite{2022MNRAS.511.3814H} showed that the 8:5 and 5:3 can be built from an original 3:2 -- 3:2 resonance if planets enter the inner cavity of the disc. Their model contains two critical assumptions. It invokes enhanced disc-driven eccentricity damping (about 50 times more efficient than expected for typical type-I migration) which was found to be advantageous for obtaining the observed Trappist-1 resonant chain.
Moreover, it involves a requirement for the timing of arrival of the outer planets to ensure that the inner system does not evolve past the 8:5 -- 5:3 ratios. Planet e must enter in resonance with planet d at the appropriate time to block the migration of planet c in a 3-body c-d-e Laplace resonance when the former is in the right orbital position; the existence and extent of the appropriate time window depend on the migration parameters introduced to mimic planet-disc interactions.
Drawing from these ideas, we develop a model whose key elements are independent of disc migration parameters, thereby removing the need for specific disc migration and damping efficiencies. Our model invokes a simple timing constraint for the arrival of the outer planets, dependent only on the rate of recession of the inner edge of the disc.\\

\begin{figure}[!tp]
\centering
\begin{tikzpicture}
    \draw (0, 0) node[inner sep=0] {\includegraphics[width=1.\textwidth]{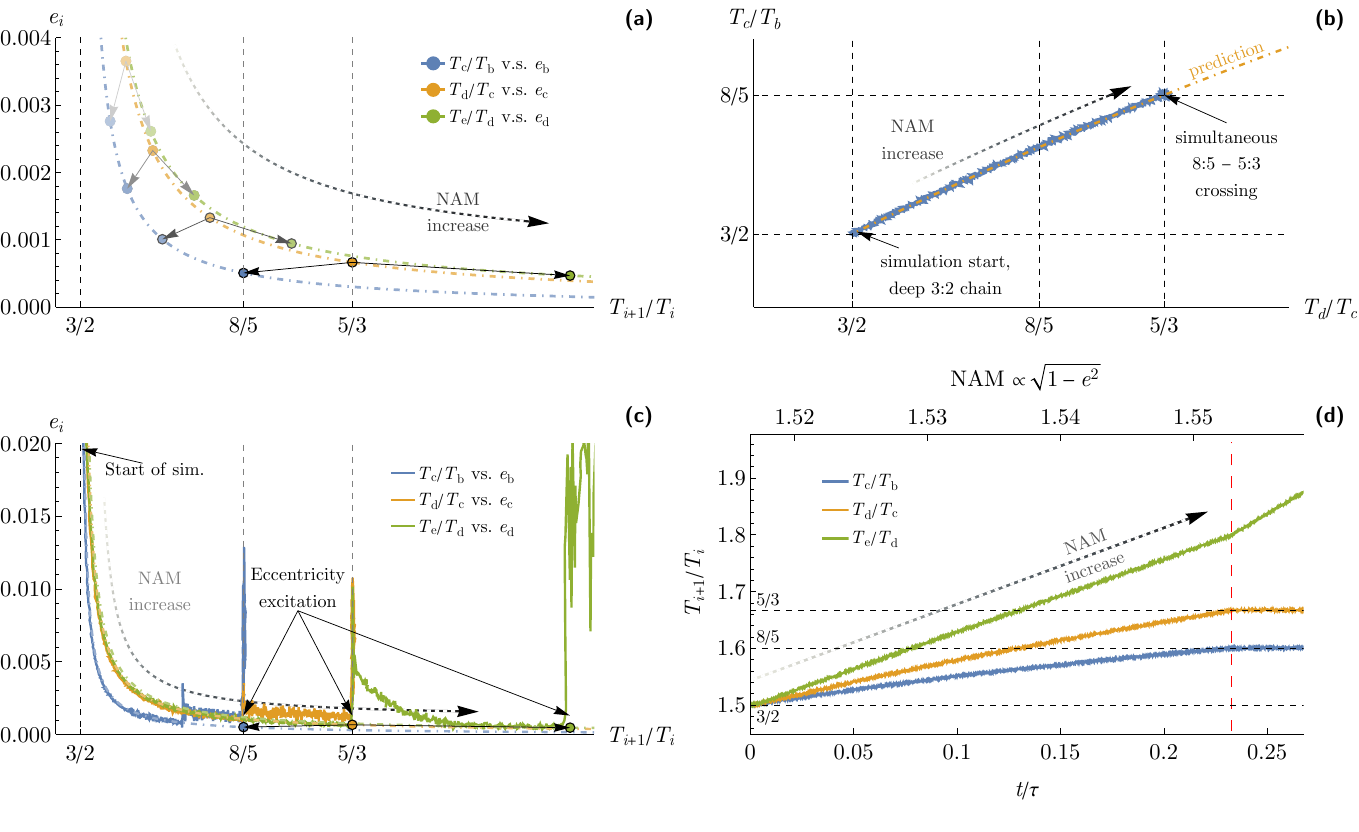}};
    \draw (-1, 5) node {\bf(a)};
    \draw (7.5, 5) node {\bf(b)};
    \draw (-1, 0.25) node {\bf(c)};
    \draw (7.5, 0.25) node {\bf(d)};
\end{tikzpicture}
\caption{{\bf Evolution of a 3:2 resonant chain under dissipation}.
{In panel a, coloured dot-dashed curves show predicted equilibria for a Trappist-1b,c,d,e system locked in a 3:2 -- 3:2 -- 3:2 resonance, in the orbital period ratio vs.\ eccentricity plane, for varying normalised angular momentum values (\emph{NAM}, the global resonant gauge of the system's state, which increases as indicated by the dotted arrow). When the planets are deep in resonance, their orbital state must lie at an equilibrium state determined solely by the NAM value. Four fixed values for the NAM are chosen for explanatory purposes, and the corresponding equilibrium period ratios and eccentricities state for the four planets are all linked together as shown by the arrows. For one of them (arrows in black), $T_{\mathrm{c}}/T_{\mathrm{b}}\simeq 8/5$ and $T_{\mathrm{d}}/T_{\mathrm{c}}\simeq 5/3$, which are the currently observed period ratios of these planets.} 
{Panel b shows this simultaneous crossing of the 8:5 – 5:3 resonance in b-c and c-d period ratios plane in a simulation of the Trappist-1b,c,d,e system starting from a deep 3:2 chain as the NAM increases in time.}
{Panel c shows this evolution in the period-rato vs.\ $e$ plane (as in panel a); lighter coloured dot-dashed lines are the same as in panel (a), while continuous lines represent the simulated evolution (some lines appear on top of each other). Notice the kick in eccentricity at the simultaneous crossing of the 8:5 -- 5:3 resonance for planets b-c-d.}
{For the same run, panel d shows with coloured continuous lines the evolution of the period ratios in time (expressed in units of a timescale $\tau$ in the bottom axis) as the NAM increases (top axis). The simultaneous 8:5 – 5:3 crossing (predicted in the top panels) is indicated by a vertical dashed red line.}}
\label{fig:1}
\end{figure}
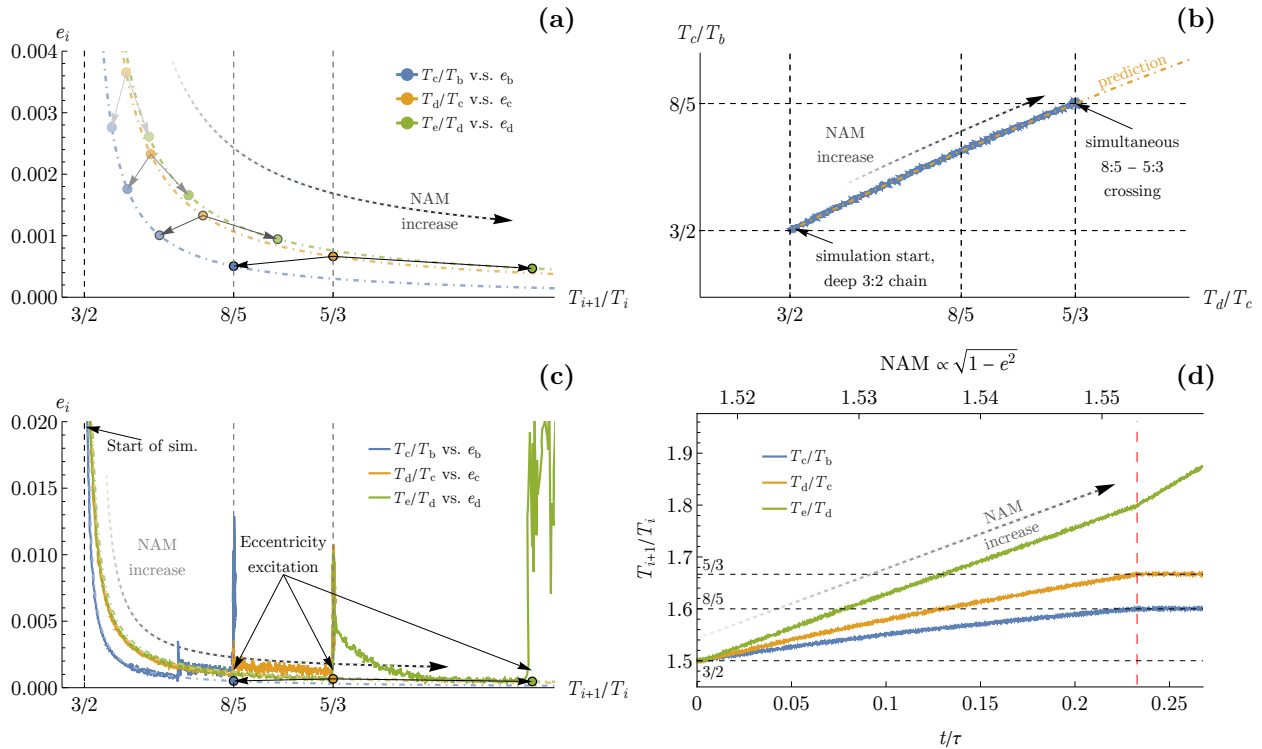

Like \cite{2018MNRAS.476.5032P,2022MNRAS.511.3814H} we consider an inner and an outer Trappist-1 sub-system.
The first step is to consider the evolution of the inner system alone near the favoured 3:2 commensurabilities: specifically how the orbital period ratios can increase so that the 8:5 -- 5:3 period ratios for planets b, c and d are reached, and how this resonant repulsion naturally corresponds to a decrease in eccentricities \cite{2012ApJ...756L..11L,2013AJ....145....1B,2019A&A...625A...7P}.
Note that in a first-order resonant chain the actual period ratios between adjacent planets are not exactly equal to the resonant ratios (e.g.\ 3:2) \cite{2013A&A...556A..28B}. Instead, these ratios depend on the planets’ eccentricities, and the latter are all correlated to each other by a global resonant gauge, namely the total angular momentum of the system normalized by the reference resonant location (\emph{NAM}; see Fig.\ \ref{fig:1}(a) for e.g.\ a 3:2 -- 3:2 -- 3:2 four-planet chain, and Methods section). Thus, period ratios and eccentricities do not evolve independently of each other: processes that result in NAM increase (such as dissipation) result in a predictable concurrent increase of all period ratios by lowering the eccentricities, associated with a rapid perihelion precession while maintaining formal libration of the resonant angles (known as resonant repulsion). It is striking that, whatever the initial NAM (i.e.\ eccentricities) was, following dissipation-driven NAM increase one finds that for a value of the NAM such that planets b and c are close to a 8:5 period ratio, planets c and d are close to the 5:3 ratio. Thus, the fact that planets b, c and d are observed to be close to these period ratios is no coincidence, but is evidence of a past evolution in the 3:2 resonant chain \cite{2018MNRAS.477.1414C,2018MNRAS.476.5032P,2022MNRAS.511.3814H}.

Fig.\ \ref{fig:1} depicts this process from quantitative grounds: originating deep in a 3:2 -- 3:2 -- 3:2 resonant chain after migration and with a specific value of the NAM dictated by their eccentricities, planets b, c, d (in the disc cavity) and e evolve under dissipation, increasing their NAM.
Fig.\ \ref{fig:1}(b,c,d) show that all period ratios increase as expected \cite{2012ApJ...756L..11L,2013AJ....145....1B,2014A&A...566A.137D}. When the period ratio $T_{\mathrm{c}}/T_{\mathrm{b}}$ reaches the value 8:5, $T_{\mathrm{d}}/T_{\mathrm{c}}$ reaches 5:3 simultaneously. As these resonances are crossed divergently, the planets' eccentricities suddenly jump to larger values (Fig.\ \ref{fig:1}(c)). If the eccentricities increase by more than a factor of two, as it happens here given the measured planetary masses (note that planets b and c are the most massive in the inner system), theory predicts that the original first-order resonant chain is broken \cite{2014A&A...566A.137D,2020MNRAS.494.4950P}. Thus, further NAM increase does not alter the $T_{\mathrm{c}}/T_{\mathrm{b}}$ and $T_{\mathrm{d}}/T_{\mathrm{c}}$ ratios any further (Fig.\ \ref{fig:1}(d)): the planets are close to, but not inside the 8:5 and 5:3 resonances because capture in these resonances is impossible during divergent migration. This scheme is a first step to a more accurate reconstruction of the dynamical history of the Trappist-1 system, and demonstrates that, although the presence of resonances indicates that migration had to have happened, other processes must have been at play concurrently with migration to shape its current architecture. If this scenario is appealing, the relevant questions are (i) what is the plausible origin of dissipation (NAM increase) and (ii) how many planets were involved in the original resonant chain?

For the specific dissipation mechanism, a first possibility is efficient dissipation onto the planets  (\cite{2012ApJ...756L..11L,2013AJ....145....1B,2014A&A...566A.137D,2018MNRAS.476.5032P,2019A&A...625A...7P}; see Method section). However, if such dissipative force had been so efficient to drive planets b, c and d to a 8:5 -- 5:3 period ratio from a primordial 3:2 -- 3:2 configuration, it would have also been able to rapidly damp their eccentricities after their jump, thus restoring the 3:2 -- 3:2 resonant chain and restarting the divergent evolution (see Extended Fig.\ \ref{fig:tide_repuls}). Another possibility is that planets b and c were closer to the star than the inner edge of the disc because either they have been pushed there by the migration of planets d and e, or they opened a gap in the disc, or the inner disc cavity expanded due to magnetic torques and photo-evaporation, or a combination of these processes. In this case, if planet d remained at the inner disc’s edge, planet c would have felt a negative one-sided Lindblad torque (OLT) from the disc (see Methods) which pushed c inwards, away from d. This is equivalent to NAM increase, due to the resonance gauge (Fig.\ \ref{fig:1}(a)). However, because this push acts directly on planet c, it would have continued after crossing the 5:3 resonance between planets d and c, bringing the system away from the observed period ratio (unless a fortuitous and timely disappearance of the disc is invoked, or the assembly of a 3-body resonance between planets c, d and e at a suitable time, as in \cite{2022MNRAS.511.3814H}).
\\

\begin{table}[h!]
\centering
\begin{tabular}{|p{1.5cm} p{5cm} p{6.25cm} p{2.2cm}|} 
 \hline
 {Phase} & {Dynamical step} & {Main driving processes} & Figure \\ [0.5ex] 
 \hline\hline
 & & & \\ 
 Phase 0 &	Assembly of primordial b-c-d-e 3:2 chain, planets b-c-d fall into inner cavity, planet e reaches the inner edge & {Classical type-I migration, \newline clearing of inner disc, \newline tidal dissipation} & Extended Data Fig.\ \ref{fig:add:bcdeBuild}\\ 
 & & & \\ 
 Phase 1 &	Planet e recedes with the inner edge and planets b-c-d divergently cross the 8:5 -- 5:3 resonance via NAM increase, OLT re-compactifies them & {Expansion of the inner edge, \newline tidal dissipation, \newline outer Lindblad torque (OLT)} & Fig.\ \ref{fig:1}\\\ 
 & & & \\ 
 Phase 2 &	Joining of the inner and outer sub-systems and assembly of the full Trappist-1 chain & {Tidal dissipation + OLT (inner system); \newline classical type-I migration (outer system); \newline expansion of the inner edge} & Fig.\ \ref{fig:2}\\ 
 & & & \\ 
 \hline
\end{tabular}
\caption{Chronological sequence of events leading to the assembly of the full Trappist-1 chain in our model.}
\label{tbl:phases}
\end{table}

{\bf Results.} Building on these ideas, we propose here a novel way to obtain the repulsion of the orbits of Trappist-1b,c,d which is largely insensitive to orbital damping parametrisations, pivoting instead on physical mechanisms that are expected to occur in the environment where planets form. We only presuppose that the inner edge of the disc receded with time\cite{2017A&A...601A..15L,2017A&A...604A...1O} while the inner system was locked in the 3:2 resonant chain. We assume a 4-planet 3:2 resonant chain like that simulated in Fig.\ \ref{fig:1}, with planet e sitting at the inner edge of the disc being our preferred scenario (this is phase 0 in Table \ref{tbl:phases}; see Supplementary material). As the disc’s edge recedes, planet e recedes in concert, as it is anchored to it by the so-called co-orbital corotation torque \cite{2008EAS....29..165M}. Consequently, the separation between planets e and d increases in time, which causes NAM increase and the increase of all period ratios, due to the resonance gauge. After the 8:5 and 5:3 resonance crossing and the break-down of the 3:2 chain, planet e keeps receding, leaving behind planets b, c and d unaffected (Fig \ref{fig:1}(d); in this simulation, the timescale $\tau$ -- describing the inner edge recession rate, $\tau = r_\mathrm{ed,0}/(\mathrm{d}r_{{\mathrm{ed}}}/\mathrm{d} t)$ -- was $700$ kyrs). Simultaneously, planet d starts to migrate inward driven by an OLT; this migration is necessarily slow since only its outer 2:1 resonance falls in the disc \cite{2022MNRAS.511.3814H}. Thus, planets c and d are now slowly convergently migrating on moderately eccentric orbits which were pumped during the earlier divergent migration and resonance crossing, and they could be captured in the 5:3 resonance. Then, they keep slowly migrating inward together, with planet c capturing b in the 8:5 resonance; eccentricity damping for planets inside the gas cavity is provided by tides inside the planets. The mysterious 8:5 -- 5:3 resonance is thus established (see phase 1 in Table \ref{tbl:phases}).

\begin{figure}[!tp]
\centering
\includegraphics[width=1.\textwidth]{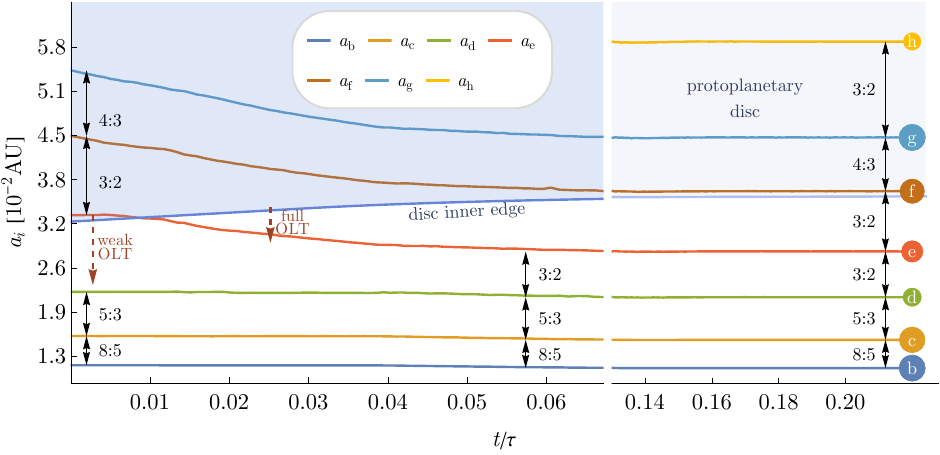}\caption{{\bf Joining of the inner and outer subsystems}.
{The evolution of the semi-major axes is shown with continuous coloured lines for all planets, which are also labelled by circles whose size reflect the observed size of the planets}. In our favoured scenario, planet e followed the inner edge while planets b, c and d were in the inner cavity. 
{Planets f, g and h join the system later, undergoing inward disc-driven migration (the disc is depicted by the top shaded area, with the lighter shading on the right indicating that the disc will eventually disperse). 
Since planet g is more massive than f and h, it is likely to capture f in resonance (the 4:3 being the favoured commensurability \cite{2022A&A...658A.170T}) and the two planets migrate in together. When f reaches the 3:2 resonance with e, the combined push from g-f is sufficient to dislodge planet e from the inner edge (shown as a blue line enclosing the top shaded area). Driven by a full OLT, the orbit of planet e decays faster than f and g, approaching the 3:2 resonance with d. Planet h is the last one to migrate in and capture g in the 3:2 resonance to complete the chain.}}
\label{fig:2}
\end{figure}

In the current Trappist-1 system planet e is in the 3:2 period ratio with planet d, not arbitrarily away from it, as Fig.\ \ref{fig:1}(d) would suggest. The most reasonable explanation is that planets f, g and h were not part of the original resonant chain, but migrated into the inner part of the disc at some later time, as expected in the ringed-formation paradigm \cite{2017A&A...604A...1O,2019A&A...627A.149S,2022NatAs...6...72M}. As planet f approached e, the two got captured in the 3:2 resonance, exciting planet e's eccentricity and dislodging it from the disc's inner edge (\cite{2008EAS....29..165M,2014MNRAS.437...96F}, see Supplementary material); in fact, planet g is more massive than f, so the f-g pair had likely already locked into the most favoured\cite{2022A&A...658A.170T} 4:3 resonance before approaching planet e, although this is not strictly necessary (see additional material). Planet e thus left the inner edge and underwent OLT-driven inward migration approaching planet d. At this moment no resonance with planet d has been re-established, so the inner system remained unaffected.

During its inward migration, planet e had a non-vanishing forced eccentricity due to the resonant interaction with f. Planet e must, however, have crossed a series of high order resonances with planet d (9:5, 5:3, 8:5), which it had to skip before reaching the 3:2 period ratio with d, where it is observed today. Capture in resonances of order higher than 1 is only a probabilistic event, while it is guaranteed for first-order resonances in the adiabatic limit. In addition, planet e experienced NAM-increase with respect to planet f due to the former's initially fast, full-OLT-driven migration (see Method section): this evolution is naturally associated with an efficient damping of the eccentricities, reducing the strength of the higher-order resonances and allowing for a smooth joining of the inner and outer chains. The advantage of this mechanism is that it relies solely on the analytical evolution tracks depicted in Fig.\ \ref{fig:1}, and not on enhanced disc-driven $e$-damping efficiencies (which may not be guaranteed \cite{2023A&A...670A.148P}, especially near the disc inner edge \cite{2021A&A...648A..69A}). 
Our experiments show that, depending on the phases of the angles at approach and with different disc profiles, in up to $22\%$ of simulations the higher-order resonances between e and d are skipped and the 8:5 -- 5:3 resonance for planets b-c-d is not disrupted, allowing for a successful joining of the inner and outer system at the required period ratios at the end of this assembly phase (see the Supplementary discussion \emph{Joining the inner and the outer systems}, Extended Data Figure \ref{fig:add:JoiningTheSystemTwoWays} and Extended Data Table \ref{tbl:Prob} for a parameter study with different surface densities). This presupposes that the inner edge has not moved outwards enough to bring the period ratio between planets d and e to be larger than 2:1 before planet f approaches the inner system, to avoid planets c and d being subsequently captured in the 2:1 resonance. A timing condition concerning the arrival of the outer planets to join the inner chain is implicit in a model that assumes a two-subsystem formation history. In our model, we also specifically require that planet f reach the inner system before the inner edge would have take planets d and e outside the 2:1 period ratio (the timing of which is set by the average inner edge recession rate after the 8:5 -- 5:3 b-c-d resonance has been crossed).
Finally, planet h, being less massive than g, would have migrated in and captured g into their 3:2 resonance at a later time, establishing the whole observed  8:5 -- 5:3 -- 3:2 -- 3:2 -- 4:3 -- 3:2 chain (Fig.\ \ref{fig:2}; see also phase 2 in Table \ref{tbl:phases}). \\

\begin{figure}[!tp]
\centering
\begin{tikzpicture}
    \draw (0, 0) node[inner sep=0] {\includegraphics[width=1.\textwidth]{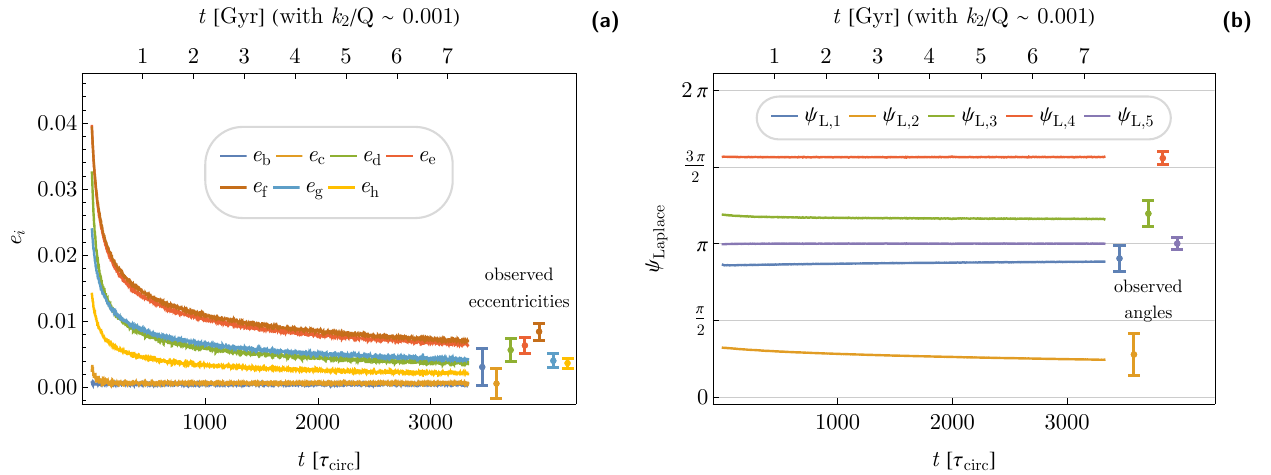}
};
    \draw (-1, 2.75) node {\bf(a)};
    \draw (7.5, 2.75) node {\bf(b)};
\end{tikzpicture}
\caption{{\bf Post-nebular evolution under tidal dissipation of the Trappist-1 system}.
Panel a shows the decay of the eccentricities due to tidal $e$-damping, showing that for reasonable tidal parameters the currently observed eccentricities are reached after around 2500 to 3000 circularisation timescales of planet b (error bars represent the 1$\sigma$ confidence limits from \cite{2021PSJ.....2....1A}). Panel b shows the libration of the (reduced) Laplace angles defined in Eq.\ \eqref{eq:LaplaceAnglesLibrating} compared with the libration centres and amplitudes (shown as error bars) observed from different draws from the posterior distribution from \cite{2021PSJ.....2....1A}.\\}
\label{fig:3}
\end{figure}

After gas removal and the disappearance of the disc, the planets undergo multi-Gyr tidal evolution. Fig.\ \ref{fig:3}(a) shows that, for realistic tidal parameters \cite{2022MNRAS.515.2373B}, the eccentricities of all planets are consistent with those determined by observations \cite{2021PSJ.....2....1A}, within the uncertainties, after about 2500 to 3000 circularisation timescales of planet b. In the resonant chain that our model produces, the planets are in Laplace resonances by triplets, as currently observed\cite{2021PSJ.....2....1A} (Fig.\ \ref{fig:3}(b)); moreover, neighbouring planet pairs between Trappist-1c and g are in two-body mean motion resonances. Planets b' and c's eccentricities reached vanishingly low values due to their tidal evolution (Fig.\ \ref{fig:3}(a)), so that their 8:5 resonance angles stopped librating and currently show circulation. Still, all planets were in two-body resonances with their neighbours at the end of the disc-phase assembly of the system, which locked their period ratios close to the currently observed ones. This explains how the mysterious high-order 8:5 -- 5:3 period ratios for planets b, c and d are not fortuitous, but indicate a past evolution where both these resonances were active. 
While the Laplace resonant states have been verified observationally \cite{2021PSJ.....2....1A}, proving the two-body resonance states requires a precise knowledge of the orbits’ longitudes of pericenter, which is difficult to achieve. Improved characterization of the dynamical state of the Trappist-1 system will eventually allow to verify our model prediction.\\

{\bf Discussion.} 

Our model builds on previous ideas such as the separation in two sub-systems \cite{2018MNRAS.476.5032P,2022MNRAS.511.3814H} and the role of the disc inner edge in anchoring planets \cite{2006ApJ...642..478M,2017A&A...604A...1O}. Here, we are able to explain for the first time the complex orbital architecture of the Trappist-1 planets in a model that is largely insensitive to specific migration and eccentricity damping efficiency parameters, by pivoting on  physical processes that are expected to occur in protoplanetary discs. 
The recession of the inner edge provides a natural dynamical pathway to assemble the otherwise unlikely(\cite{2022A&A...658A.170T}) 8:5 -- 5:3 resonances. 
Such inner edge evolution has been predicted by theoretical models \cite{2017A&A...601A..15L}, but is so far inaccessible to observations. Our work provides indirect evidence of its role in the assembly of the Trappist-1 system. In our model, the (unknown) rate of the inner edge recession rate after the 8:5 -- 5:3 b-c-d resonance crossing should not be too fast that planets d-e exceed the 2:1 period ratio before the arrival time (also unknown) of planet f (see Supplementary discussion). However, we note that the inner edge location is set by the magnetic truncation radius $R_\mathrm{t}\propto \dot{M}_\mathrm{g}^{-2/7}$ \cite{2017A&A...601A..15L}, with the gas accretion rate $\dot{M}_\mathrm{g}\propto(1+t/\tau_\mathrm{acc})^{-\eta}$, where $\tau_\mathrm{acc}$ is a (time-averaged) accretion timescale \cite{2013ApJ...778..169B}, and $\eta\gtrsim 1$, as expected for viscous accretion \cite{2023ASPC..534..539M}. Thus, the inner edge recession rate is expected to slow down over time.
Moreover, our model shows that the strong eccentricity damping needed to obtain these specific configurations can be the result of divergent migration in a resonant configuration inside the disc cavity, rather than direct damping from the disc, thus removing a tension with the hydrodynamical simulations \cite{2022MNRAS.511.3814H}.
All these ingredients, essential to reproduce the Trappist-1 system within our current understanding of planet-disc interactions, may have played an important role in other planetary systems as well and open the possibility of a precise reconstruction of their dynamical history as we have done here for the Trappist-1 system.\\

\textbf{Methods}\\

\ul{Mean motion resonances and Laplace angles}\\
The analytical treatment of chains of mean motion resonances is the subject of many works \cite{2013A&A...556A..28B,2017A&A...605A..96D,2018CeMDA.130...54P} and is reasonably well understood, so we do not reproduce it here. We only detail the main points that are used in the analytical calculations.

Consider $N$ planets labelled by their distance from the star. If planets $i$, $i+1$ are close to a $p_i+q_i$:$p_i$ mean motion resonance, with $p_i$ and $q_i$ integers, then the period ratio $T_{i+1}/T_{i}\simeq (p_i+q_i)/p_i$. After expanding the gravitational potential describing planet-planet interactions in a Fourier series, all terms of the potential that do not contain the resonant combination $(p_i+q_i)\lambda_{i+1} - p_i \lambda_i$ can be averaged out (dropped) since they are fast angles; moreover, by the d'Alembert rules, the only harmonics that appear at lowest order in the eccentricities are
\begin{equation}\label{eq:ResonantAngles}
\begin{split}
    &\text{$\psi_{i}^{(q_i,0)}:=(p_i+q_i)\lambda_{i+1} - p_i \lambda_i - q_i\varpi_i$ (associated with a term proportional to $e_i^{q_i}$),}\\
    &\text{$\psi_{i}^{(q_i-1,1)}:=(p_i+q_i)\lambda_{i+1} - p_i \lambda_i - (q_i-1)\varpi_i-\varpi_{i+1}$ (term proportional to $e_i^{q_i-1}e_{i+1}^{1}$),}\\
    &\text{$\psi_{i}^{(q_i-2,2)}:=(p_i+q_i)\lambda_{i+1} - p_i \lambda_i - (q_i-2)\varpi_i-2\varpi_{i+1}$ (term proportional to $e_i^{q_i-2}e_{i+1}^{2}$),}\\
    &\text{...,}\\
    &\text{$\psi_{i}^{(0,q_i)}:=(p_i+q_i)\lambda_{i+1} - p_i \lambda_i - q_i\varpi_{i+1}$ (term proportional to $e_{i+1}^{q_i}$)}.
\end{split}
\end{equation}
The $\psi_{i}$ are called the (2-body) resonant angles for the $p_i+q_i$:$p_i$ mean motion resonance and represent the relevant interaction terms that drive the dynamics. For example, for the 3:2 mean motion resonance between planets Trappist1-d and e, $p=2$ and $q=1$ (first order resonance) and there are two resonant angles: $3{\lambda_{\mathrm{e}}} - 2{\lambda_{\mathrm{d}}}-{\varpi_{\mathrm{d}}}$ and $3{\lambda_{\mathrm{e}}} - 2{\lambda_{\mathrm{d}}}-{\varpi_{\mathrm{e}}}$; for the 5:3 mean motion resonance between planets Trappist1-c and d, $p=3$ and $q=2$ (second order resonance) and there are three resonant angles: $5{\lambda_{\mathrm{d}}} - 3{\lambda_{\mathrm{c}}}-2{\varpi_{\mathrm{c}}}$, $5{\lambda_{\mathrm{d}}} - 3{\lambda_{\mathrm{c}}}-{\varpi_{\mathrm{c}}}-{\varpi_{\mathrm{d}}}$ and $5{\lambda_{\mathrm{d}}} - 3{\lambda_{\mathrm{c}}}-2{\varpi_{\mathrm{d}}}$.
When two planets are in the resonance, some or all resonant angles are observed to be librating, representing a dynamical state contained within a resonant island in phase space. Note that all resonant interaction terms are proportional to the eccentricities to the power $q_i$, and they therefore contribute less for larger $q_i$'s or smaller eccentricities (i.e.,\ the resonant islands cover less volume in phase space).

We note that if a triplet of consecutive planets $i$, $i+1$, $i+2$ lies in a resonant chain, one can build appropriate combinations of the 2-body resonant angles of the two pairs to obtain an angle that depends on the mean longitudes of the three planets but not on any of the pericentres. For example, with a $p_i+1$:$p_i$ resonance for the inner pair and a $p_{i+1}+1$:$p_{i+1}$ for the outer pair, then the combination of the resonant angles $(p_i+1)\lambda_{i+1} - p_i \lambda_i - \varpi_{i+1}$ (inner pair's resonance, pericentre of the outer planet) and $(p_{i+1}+1)\lambda_{i+2} - p_{i+1} \lambda_{i+1} - \varpi_{i+1}$ (outer pair's resonance, pericentre of the inner planet) removes the dependence on the pericentre $\varpi_{i+1}$:
\begin{equation}
\begin{split}
    \psi_{i}^{(p_{i+1}+1,- (p_{i+1}+p_i+1),p_i)}  &:=\big((p_{i+1}+1)\lambda_{i+2} - p_{i+1} \lambda_{i+1} - \varpi_{i+1}\big) - \big((p_i+1)\lambda_{i+1} - p_i \lambda_i - \varpi_{i+1}\big)\\
                &=(p_{i+1}+1)\lambda_{i+2} - (p_{i+1}+p_i+1)\lambda_{i+1} + p_i \lambda_i.
    \end{split}
\end{equation}
Such 3-body angles are called Laplace angle. If both resonant angles are librating, also the Laplace angle must librate. Since the Laplace angles for triplets of planets do not involve the pericentres, but only the mean longitudes $\lambda_i$ which are determined precisely by the transit times, their libration is easier to observe compared to the libration of the 2-body resonance angles in exoplanetary systems, where information on the pericentres is harder to obtain. \\

The Laplace angles that are reported to be librating in the Trappist-1 chain by \cite{2021PSJ.....2....1A} are 
\begin{equation}\label{eq:LaplaceAnglesLibrating}
\begin{split}
\psi_{\mathrm{L},1} = 2{\lambda_{\mathrm{b}}} - 5{\lambda_{\mathrm{c}}} + 3{\lambda_{\mathrm{d}}} &= -(3{\lambda_{\mathrm{c}}} - 2{\lambda_{\mathrm{b}}}-{\varpi_{\mathrm{c}}}) + (3{\lambda_{\mathrm{d}}} - 2{\lambda_{\mathrm{c}}}-{\varpi_{\mathrm{c}}}),\\
\psi_{\mathrm{L},2} = 1{\lambda_{\mathrm{c}}} - 3{\lambda_{\mathrm{d}}} + 2{\lambda_{\mathrm{e}}} &= [-(5{\lambda_{\mathrm{d}}}-3{\lambda_{\mathrm{c}}}-2{\varpi_{\mathrm{d}}}) + 2(3{\lambda_{\mathrm{e}}} - 2{\lambda_{\mathrm{d}}}-{\varpi_{\mathrm{d}}})]/3,\\
\psi_{\mathrm{L},3} = 2{\lambda_{\mathrm{d}}} - 5{\lambda_{\mathrm{e}}} + 3{\lambda_{\mathrm{f}}} &= -(3{\lambda_{\mathrm{e}}} - 2{\lambda_{\mathrm{d}}}-{\varpi_{\mathrm{e}}}) + (3{\lambda_{\mathrm{f}}} - 2{\lambda_{\mathrm{e}}}-{\varpi_{\mathrm{e}}}),\\
\psi_{\mathrm{L},4} = 1{\lambda_{\mathrm{e}}} - 3{\lambda_{\mathrm{f}}} + 2{\lambda_{\mathrm{g}}} &= [-(3{\lambda_{\mathrm{f}}}-2{\lambda_{\mathrm{e}}}-{\varpi_{\mathrm{f}}}) + (4{\lambda_{\mathrm{g}}} - 3{\lambda_{\mathrm{f}}}-{\varpi_{\mathrm{f}}})]/2,\\
\psi_{\mathrm{L},5} = 1{\lambda_{\mathrm{f}}} - 2{\lambda_{\mathrm{g}}} + 1{\lambda_{\mathrm{h}}} &= [-(4{\lambda_{\mathrm{g}}}-3{\lambda_{\mathrm{f}}}-2{\varpi_{\mathrm{g}}}) + (3{\lambda_{\mathrm{h}}} - 2{\lambda_{\mathrm{g}}}-{\varpi_{\mathrm{g}}})]/3.
\end{split}
\end{equation}
Note that some of these angles are obtained from two-body resonant angles after dividing by an integer (the greatest common divisor of the integers multiplying the angles) \cite{2021AJ....161..290S}. However, only non-reduced angles actually appear in the Fourier expansion of the gravitational potential, while the division by an integer is equivalent to a phase-folding of the angles.\\

\ul{Resonant equilibria for a 3:2 resonance chain}\\
If all planet pairs in a planetary system are in $p_i+1$:$p_1$ 2-body resonances, planet-planet interaction terms that do not involve resonant angles can be dropped. Thus, the resonant model will not depend on any one of the longitudes $\lambda_i$, but only on their resonant combinations. This introduces, in a regime without external dissipation or forcing, an additional constant of motion $\mathcal{K}$, a resonant scaling parameter \cite{2008MNRAS.387..747M,2013A&A...556A..28B}. For an $N$-planet 3:2 chain, this constant of motion takes the form \cite{2019A&A...625A...7P}
\begin{equation}
    \mathcal{K} = \sum_{i=1}^N \left(\frac{2}{3}\right)^{i-1} m_i \sqrt{\mu_0 a_i},
\end{equation}
where $m_i$ are the planets' masses and $\mu_0 = {\mathcal{G}} M_*$ is the standard gravitational parameter of the central star.
We define the normalised angular momentum (NAM) as $\mathcal{L}/\mathcal{K}$,
where $\mathcal{L} = \sum_{i=1}^N m_i \sqrt{\mu_0 a_i (1-e_i^2)}$ is the total (orbital) angular momentum, which is itself a conserved quantity in the non-dissipative regime.
Since the planetary problem is scale-free and $\mathcal{K}$ is proportional to $a^{1/2}$, this is equivalent to reasoning in terms of semi-major axis ratios rather than absolute semi-major axes themselves. The global state of the resonant system in rescaled quantities is thus completely determined by the value of the NAM. 

With this scheme at hand, using the analytical treatment of resonances (e.g.\ \cite{2019A&A...625A...7P}), one can find, for a given value of the NAM, a resonant equilibrium in phase space by imposing libration of the resonant angles. This results in curves parametrised by the NAM value which track the resonant states as a function of the orbital elements, as shown in Fig.\ \ref{fig:1}. The fact that the resonant locations do not coincide with period ratios exactly equal to the nominal ratios $(p_i+1)/p_i$, but deviate away from exact commensurability for small eccentricities, is simple to understand: at the centre of the resonant island, the libration of a resonant angle like $(p_i+1) \lambda_{i+1} - p_i \lambda_i - \varpi$ (where $\varpi$ is either $\varpi_i$ or $\varpi_{i+1}$) imposes that $(p_i+1) \Omega_{{\mathrm{Kep}},i+1} -p_i \Omega_{{\mathrm{Kep}},i} = \frac{\D{}}{\D t}\big[(p_i+1) \lambda_{i+1} - p_i \lambda_i\big] = \frac{\D\varpi}{\D t}=:\dot\varpi$. Since $\dot\varpi$ is proportional to $1/e$, for faster perihelion precession rates (lower eccentricities) the orbital periods of the planets will be farther away from exact commensurability. Fig.\ \ref{fig:1}(b) shows with a dot-dashed orange line the same curve but in the period ratio space $T_{\mathrm{c}}/T_{\mathrm{b}}$ vs.\ $T_{\mathrm{d}}/T_{\mathrm{c}}$, showing how starting very close to a 3:2 -- 3:2 period ratio commensurability one naturally crosses the 8:5 -- 5:3 for a system undergoing NAM increase. This peculiarity of the phase space near a 3:2 -- 3:2 was already observed e.g.\ in \cite{2018MNRAS.477.1414C}. \\

\ul{Planet-disc interactions, disc structure and modeling of the inner edge}\\
To model planet-disc interactions, we follow the prescription of \cite{2008A&A...482..677C} derived from 3D hydro-dynamical simulations for planets with planet-to-star mass ratios similar to the Trappist-1 system. This prescribes the torque and eccentricity damping felt by a planet deeply embedded in the disc \cite{2002ApJ...565.1257T}, similarly to other works on the Trappist-1 system \cite{2022A&A...658A.170T,2022MNRAS.511.3814H,2022MNRAS.515.2373B}: 
\begin{equation}
\begin{split}
\Gamma_{\mathrm{tI}} = \left(\frac{\D {\mathcal{L}}}{\D t}\right)_{\mathrm{tI}} &= -\frac{{\mathcal{L}}}{\tau_{{\mathrm{mig}},{\mathrm{tI}}}},\\\left(\frac{\D e}{\D t}\right)_{\mathrm{tI}} &= -\frac{e}{\tau_{e,{\mathrm{tI}}}}.
\end{split}
\end{equation}
Like \cite{2022A&A...658A.170T,2022MNRAS.515.2373B}, we use the migration prescription from \cite{2008A&A...482..677C}, which is consistent with 3D hydrodynamical simulations of planets of this mass range embedded in discs.
We discuss in Supplementary discussion section \emph{Calculation of one-sided Lindblad Torques} the modelling of the one-sided Lindblad torque felt by a planet that has fallen into the cavity.

We model the static disc surface density as a power law multiplied by a factor to reproduce the drop in gas density at the inner cavity:
\begin{equation}\label{eq:add:InnerDiscStruct}
    \Sigma(r) = \mathcal{R} \Sigma_0 \left(\frac{r}{r_0}\right)^{-{\alpha_{\Sigma,0}}},
\end{equation}
where $\Sigma_0$ is a reference density at a reference radius $r_0$ (we take for example $r_0 = 0.025 \mathrm{AU}$, in between the current locations of planets d and e) and ${\alpha_{\Sigma,0}}$ is the slope of the un-broken power law disc (at larger separations away from the inner edge). We chose an underlying power-law exponent ${\alpha_{\Sigma,0}}=3/5$, which is appropriate for the viscously-heated region of the disc, as was done in \cite{2022MNRAS.515.2373B}. For the surface density $\Sigma_0$, we tested values spanning different orders of magnitude (see Extended Data Table \ref{tbl:Prob}).
We use a factor $\mathcal{R}$ given by
\begin{equation}\label{eq:MathcalR}
    \mathcal{R} = \mathcal{R}(r;r_{\mathrm{ed}},h_{\mathrm{ed}},w_{\mathrm{ed}}) = G\left(\frac{r-r_{\mathrm{ed}}}{h_{\mathrm{ed}} r_{\mathrm{ed}} w_{\mathrm{ed}}}\right),
\end{equation}
where $r_{\mathrm{ed}}$ is the reference location of the inner edge, $h_{\mathrm{ed}}:=h(r_{\mathrm{ed}})$ is the aspect ratio of the disc at $r_{\mathrm{ed}}$ and $w_{\mathrm{ed}}$ represents an inner edge width, which we take to be 1, although we also experimented with a value of 0.5. 
We consider a constant aspect ratio across the disc for simplicity, $h(r) = h_0$, and thus the flaring index ${{\beta_\mathrm{fl}}}=0$, like in previous works\cite{2022MNRAS.511.3814H,2022MNRAS.515.2373B}. We choose $h_0=0.05$, but, as we note in the Supplementary material, the actual value of the aspect ratio has the sole effect of re-scaling the capture eccentricities attained after the establishment of the 3:2 chain, as $e_{\mathrm{capt}}\propto h$; however, after the resonant repulsion driven by the receding inner edge, this information is completely lost.
In this work the function $G$ is given by
\begin{equation}
    G(\xi):= \frac{\tanh(2\xi)+1}{2},
\end{equation}
which satisfies $G(\xi)\approx 0$ if $\xi\lesssim -1$ and $G(\xi)\approx 1$ if $\xi\gtrsim 1$. The choice of the hyperbolic tangent is informed by non-ideal magneto-hydrodynamical simulations \cite{2017ApJ...835..230F} and has been used in many $N$-body works \cite{2017MNRAS.470.1750I,2021A&A...650A.152I}.  The specific functional form differs however slightly from similar description of discs inner edges in previous papers \cite{2017MNRAS.470.1750I,2021A&A...650A.152I}, as it has the advantage of being differentiable at all radii. This means that the local surface density slope
\begin{equation}
    {\alpha_\Sigma} = {\alpha_\Sigma}(r) := -\frac{\D\log\Sigma}{\D\log r}(r)
\end{equation}
is a continuous function such that ${\alpha_\Sigma}(r)\approx{\alpha_{\Sigma,0}}$ for $r\gtrsim r_{\mathrm{ed}}$. 
The continuity of the local slope ${\alpha_\Sigma}(r)$ is an advantage with respect to previous prescriptions when calculating the torque felt by a planet at the inner edge, which depends on ${\alpha_\Sigma}$, as it eliminates the presence of unphysical discontinuities in the torque.

Once the inner b-c-d-e 3:2 -- 3:2 -- 3:2 chain is built, we assume that planets b, c, and d fall into the inner cavity, while planet e remains in the disc and reaches the inner disc edge. We mimic this in our $N$-body simulations as an evolution where the gas around planets b, c and d is slowly removed (a clearing the inner region of the disc). To this effect, we implement a time-dependent gas surface density prescription given by 
\begin{equation}\label{eq:add:InnerDiscRemoval1}
    \Sigma(r,t) = \tilde{\mathcal{R}}(t) \Sigma_0 \left(\frac{r}{r_0}\right)^{-{\alpha_{\Sigma,0}}}.
\end{equation}
The new time-dependent $\tilde{\mathcal{R}}$ is taken here as
\begin{equation}\label{eq:add:InnerDiscRemoval2}
    \tilde{\mathcal{R}}(t) := \mathcal{R}_{\mathrm{in}} + (\mathcal{R}_{\mathrm{fin}} - \mathcal{R}_{\mathrm{in}})\tilde{G}\left(\frac{t-t_0}{t_{\mathrm{in}} - t_0}\right),
\end{equation}
where
$\mathcal{R}_{\mathrm{in}} = \mathcal{R}(r;r_{{\mathrm{ed}},{\mathrm{in}}},h_{{\mathrm{ed}},{\mathrm{in}}},w_{{\mathrm{ed}},{\mathrm{in}}})$, $\mathcal{R}_{\mathrm{fin}} = \mathcal{R}(r;r_{{\mathrm{ed}},{\mathrm{fin}}},h_{{\mathrm{ed}},{\mathrm{fin}}},w_{{\mathrm{ed}},{\mathrm{fin}}})$, and $\tilde{G}(\xi)$ is a function that is zero for $\xi<0$, 1 for $\xi>1$ and smooth for $0<\xi<1$; we take for example
\begin{equation}\label{eq:tildeG}
    \tilde{G}(\xi) := \begin{cases}
          0 \quad &\text{if } \, \xi \leq 0, \\
          \large(\cos(\pi (\xi+1))+1\large)/2 \quad &\text{if } \, 0 < \xi < 1, \\
          1 \quad &\text{if } \, \xi \geq 1.
     \end{cases}
\end{equation}
The effect of equations \eqref{eq:add:InnerDiscRemoval1} through \eqref{eq:tildeG} is to simulate the clearing of the inner region of the disc so that the final inner edge position is at $r_{{\mathrm{ed}},{\mathrm{fin}}}$, which we assume lies between the orbits of planets d and e. The exact functional form is unimportant, as is the timescale over which this evolution takes place ($t_\mathrm{in}-t_0$), as long as the evolution is adiabatic (see also the Supplementary discussion section \emph{Evolution of the Trappist-1b,c,d,e inner system: capture in the 3:2 chain and clearing of the inner disc)}.

The shift in the inner edge position is simply accomplished by changing the value of $r_{\mathrm{ed}}$ in \eqref{eq:MathcalR}. We experimented with various functional forms for the drift of the inner edge, which all gave similar results based on the theoretical tracks depicted in Fig. \ref{fig:1}.
At the end of the assembly of the Trappist-1 system, the disc is removed by multiplying the surface density by an exponentially decreasing function of time.\\

\ul{Tidal dissipation}\\
As the Trappist-1 planets are very close to the central star, they are expected to experience significant tidal dissipation \cite{1966Icar....5..375G,2013AJ....145....1B}. In our simulations, tidal effects act as eccentricity-damping dissipative forces with
\begin{equation}
    \left(\frac{\D e}{\D t}\right)_{\mathrm{tide}} = -e \frac{21 \Omega_{\mathrm{Kep}}}{2} \frac{k_2}{Q} \frac{M_*}{m_{\mathrm{pl}}} \left(\frac{R}{a}\right)^5 = : -\frac{e}{\tau_{e,{\mathrm{tide}}}},
\end{equation}
where $R$ is the planet's radius, $Q$ is the tidal quality factor and $k_2$ is the planetary Love number \cite{1966Icar....5..375G}. Tides inside the star are neglected. There are large uncertainties on the value of $k_2/Q$, but this can be estimated to be of the order of $10^{-3}$ from dynamical arguments \cite{2018MNRAS.476.5032P,2022MNRAS.511.3814H} or interior modelling \cite{2022MNRAS.515.2373B}. Note that the role of eccentricity damping in the assembly phase is to counteract the push provided by the torque \cite{2018CeMDA.130...54P} (in this case the outer Lindblad torque). Since the torque is proportional to the local surface density around the inner edge of the disc, any uncertainty in the value of $k_2/Q$ can be carried over to a modified local surface density.\\

\textbf{Data availability}\\
The time series of the simulations displayed in the manuscript, and data to plot the analytical curves of Fig.\ \ref{fig:1}, are available at \url{https://github.com/GabrielePichierri/FormingTrappist-1}.
Data corresponding to the observed physical and orbital state of the Trappist-1 system are taken from \url{https://github.com/ericagol/TRAPPIST1_Spitzer}.\\

\textbf{Code availability}\\
The N-body integrations were run using the publicly available SWIFT subroutine package (\url{https://www.boulder.swri.edu/~hal/swift.html}), which has been modified to add the needed additional forces. The additional subroutines are available upon request from the corresponding author (G.P.). The analytical calculations have been performed using the computational software program Mathematica.\\

\textbf{Acknowledgments}\\
The authors are grateful to the anonymous referees for their comments which significantly improved the clarity and quality of the manuscript.

G.P.\ is grateful for support from the European Research Council Starting Grant 757448-PAMDORA and from the Barr Foundation for their financial support.
A.M.\ is grateful for support from the ERC advanced grant HolyEarth N. 101019380 and to Caltech for the visiting professor program that he could benefit from. 
K.B.\ is grateful to Caltech’s center for comparative planetology, the David and Lucile Packard Foundation, and the National Science
Foundation (grant number: AST 2109276) for their generous support. This study is supported by the Research Council of Norway through its Centres of Excellence funding scheme, project No. 332523 PHAB (R.B.).
G.P.\ thanks Max Goldberg and Bertram Bitsch for helpful discussions.\\

\textbf{Author Contribution Statement}\\
G.P.\ conceived the model, designed the $N$-body simulations, modified the $N$-body code with the additional routines to model the additional effects (migration, the inner edge of the disc and its evolution, and the OLT implementation), performed the numerical experiments, analysed the data, produced the figures and wrote the paper.
A.M.\ conceived and designed the model, suggested to use resonant repulsion as a mechanism to circularise the orbits at the inner edge, supplied part of the code to model migration, and wrote the paper.
K.B.\ contributed the part of the $N$-body code to model the planetary tides, helped in designing the $N$-body simulations and in writing the paper.
R.B.\ originated the discussion on the formation of the Trappist-1 system, contributed data on the planets' physical and orbital properties and helped in writing the paper.\\

\textbf{Competing Interests Statements}\\
The authors declare no competing interests.\\

\newcommand{\actaa}{Acta Astron.}   
\newcommand{\araa}{Annu. Rev. Astron. Astrophys.}   
\newcommand{\areps}{Annu. Rev. Earth Planet. Sci.} 
\newcommand{\aar}{Astron. Astrophys. Rev.} 
\newcommand{\ab}{Astrobiology}    
\newcommand{\aj}{Astron. J.}   
\newcommand{\ac}{Astron. Comput.} 
\newcommand{\apart}{Astropart. Phys.} 
\newcommand{\apj}{Astrophys. J.}   
\newcommand{\apjl}{Astrophys. J. Lett.}   
\newcommand{\apjs}{Astrophys. J. Suppl. Ser.}   
\newcommand{\ao}{Appl. Opt.}   
\newcommand{\apss}{Astrophys. Space Sci.}   
\newcommand{\aap}{Astron. Astrophys.}   
\newcommand{\aapr}{Astron. Astrophys. Rev.}   
\newcommand{\aaps}{Astron. Astrophys. Suppl.}   
\newcommand{\baas}{Bull. Am. Astron. Soc.}   
\newcommand{\caa}{Chin. Astron. Astrophys.}   
\newcommand{\cjaa}{Chin. J. Astron. Astrophys.}   
\newcommand{\cqg}{Class. Quantum Gravity}    
\newcommand{\epsl}{Earth Planet. Sci. Lett.}    
\newcommand{\expa}{Exp. Astron.}    
\newcommand{\frass}{Front. Astron. Space Sci.}    
\newcommand{\gal}{Galaxies}    
\newcommand{\gca}{Geochim. Cosmochim. Acta}   
\newcommand{\grl}{Geophys. Res. Lett.}   
\newcommand{\icarus}{Icarus}   
\newcommand{\ija}{Int. J. Astrobiol.} 
\newcommand{\jatis}{J. Astron. Telesc. Instrum. Syst.}  
\newcommand{\jcap}{J. Cosmol. Astropart. Phys.}   
\newcommand{\jgr}{J. Geophys. Res.}   
\newcommand{\jgrp}{J. Geophys. Res.: Planets}    
\newcommand{\jqsrt}{J. Quant. Spectrosc. Radiat. Transf.} 
\newcommand{\lrca}{Living Rev. Comput. Astrophys.}    
\newcommand{\lrr}{Living Rev. Relativ.}    
\newcommand{\lrsp}{Living Rev. Sol. Phys.}    
\newcommand{\memsai}{Mem. Soc. Astron. Italiana}   
\newcommand{\maps}{Meteorit. Planet. Sci.} 
\newcommand{\mnras}{Mon. Not. R. Astron. Soc.}   
\newcommand{\nat}{Nature} 
\newcommand{\nastro}{Nat. Astron.} 
\newcommand{\ncomms}{Nat. Commun.} 
\newcommand{\ngeo}{Nat. Geosci.} 
\newcommand{\nphys}{Nat. Phys.} 
\newcommand{\na}{New Astron.}   
\newcommand{\nar}{New Astron. Rev.}   
\newcommand{\physrep}{Phys. Rep.}   
\newcommand{\pra}{Phys. Rev. A}   
\newcommand{\prb}{Phys. Rev. B}   
\newcommand{\prc}{Phys. Rev. C}   
\newcommand{\prd}{Phys. Rev. D}   
\newcommand{\pre}{Phys. Rev. E}   
\newcommand{\prl}{Phys. Rev. Lett.}   
\newcommand{\psj}{Planet. Sci. J.}   
\newcommand{\planss}{Planet. Space Sci.}   
\newcommand{\pnas}{Proc. Natl Acad. Sci. USA}   
\newcommand{\procspie}{Proc. SPIE}   
\newcommand{\pasa}{Publ. Astron. Soc. Aust.}   
\newcommand{\pasj}{Publ. Astron. Soc. Jpn}   
\newcommand{\pasp}{Publ. Astron. Soc. Pac.}   
\newcommand{\raa}{Res. Astron. Astrophys.} 
\newcommand{\rmxaa}{Rev. Mexicana Astron. Astrofis.}   
\newcommand{\sci}{Science} 
\newcommand{\sciadv}{Sci. Adv.} 
\newcommand{\solphys}{Sol. Phys.}   
\newcommand{\sovast}{Soviet Astron.}   
\newcommand{\ssr}{Space Sci. Rev.}   
\newcommand{\uni}{Universe} 

\newcommand{\celmech}{Cel. Mech. Dyn. Astron.} 

\newpage

\setcounter{figure}{0}
\renewcommand{\figurename}{Extended Data Figure}
\setcounter{table}{0}
\renewcommand{\tablename}{Extended Data Table}

\begin{figure}[!h]
\centering
\begin{tikzpicture}
    \draw (0, 0) node[inner sep=0] {\includegraphics[width=1.\textwidth]{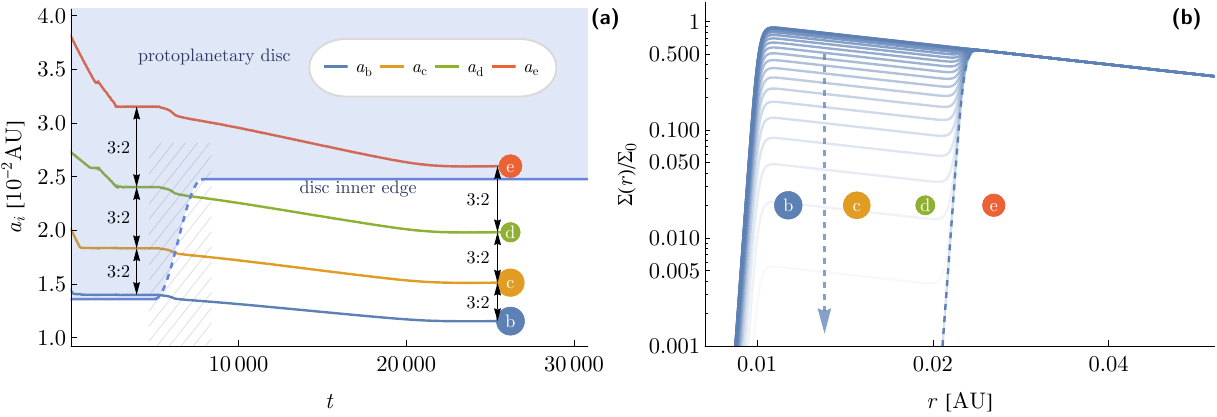}
};
    \draw (0, 3) node {\bf(a)};
    \draw (7.5, 3) node {\bf(b)};
\end{tikzpicture}
\caption{{\bf Assembly of a primordial 3:2 chain among planets Trappist-1b,c,d,e}. 
Planets b,c,d,e form a 3:2 -- 3:2 -- 3:2 chain inside the disc (top shade area), followed by planets b, c and d entering the inner disc cavity, with planet e migrating and reaching the inner edge (the blue line enclosing the top shaded area). The evolution of the semi-major axes is plotted in panel (a). Each planet is indicated by a coloured circle whose size reflects the observed size of the planet. This $N$-body simulation mimics the entry of planets b, c and d inside the inner cavity as a removal of the inner portion of the disc surrounding these planets. The corresponding evolution of the surface density in this phase is sketched in the panel (b), where the arrow indicates the drop of surface density in time (see also the dashed region in panel (a) and the shift in the initial and final position of the inner edge).}
\label{fig:add:bcdeBuild}
\end{figure}

\begin{figure}[!h]
\centering
\begin{tikzpicture}
    \draw (0, 0) node[inner sep=0] {\includegraphics[width=1.\textwidth]{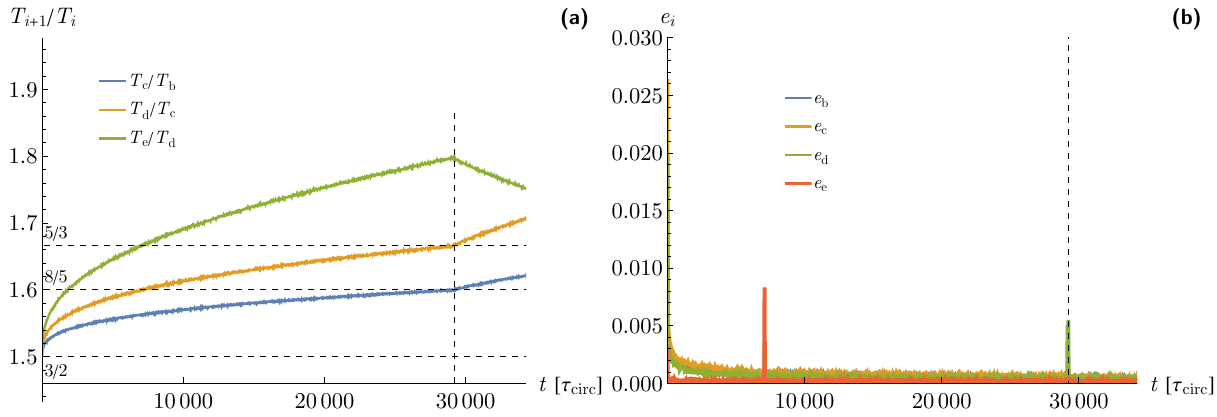}
};
    \draw (-1, 2.75) node {\bf(a)};
    \draw (7.5, 2.75) node {\bf(b)};
\end{tikzpicture}
\caption{{\bf Purely tidal evolution of a Trappist-1b,c,d,e 3:2 -- 3:2 -- 3:2 chain}. This evolution is similar to the one in Fig.\ \ref{fig:1}, with the sole difference that NAM increase is instead provided by a dissipative force onto the planets inside the cavity. The evolution of period ratios (panel (a)) and eccentricities (panel (b)) is equivalent to that of Fig.\ \ref{fig:1} as expected, until the crossing of the double 8:5 -- 5:3 resonance. At this point, unlike the case of NAM increase provided by a receding inner edge (Fig.\ \ref{fig:1}), the dissipative force quickly re-establishes a 3:2 -- 3:2 commensurability between planets b,c and d by efficiently lowering their eccentricities, thus restoring the resonant repulsion mechanism. Thus, $T_{\mathrm{c}}/T_{\mathrm{b}}$ and $T_{\mathrm{d}}/T_{\mathrm{c}}$ continue to grow past the observed 8:5 -- 5:3 ratios, while planet e jumps out of resonance. This shows that direct dissipation onto the planets alone is not as robust a mechanism to explain the assembly of the inner-most 8:5 -- 5:3 chain}
\label{fig:tide_repuls}
\end{figure}

\begin{figure}[!h]
\centering
\begin{tikzpicture}
    \draw (0, 0) node[inner sep=0] {\includegraphics[width=1.\textwidth]{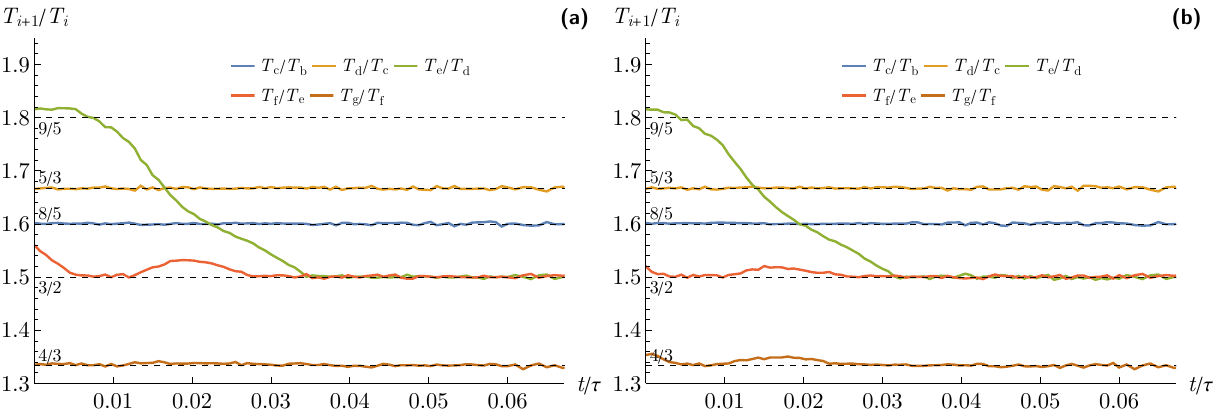}
};
    \draw (-1, 2.75) node {\bf(a)};
    \draw (7.5, 2.75) node {\bf(b)};
\end{tikzpicture}
\caption{{\bf The planets' period ratios while the inner and outer subsystems are joined.} Two examples of the joining of the inner system (with b, c and d in their 8:5 -- 5:3 resonance) with the outer system are shown (see also the left panel of Fig.\ \ref{fig:2}). Panel (a): planets f and g already in their 4:3 resonance. Panel (b): planets f and g close, but not yet inside, their 4:3 resonance. The evolution in both cases is very similar: planet e interacts with planet f via their 3:2 resonance (red curve) and starts to migrate inward with respect to planet d (green curve). Note in particular that, when planet e crosses high-order resonances with planet d, the period ratio $T_{\mathrm{f}}/T_{\mathrm{e}}$ increases slightly, which is associated with very efficient $e_{\mathrm{e}}$-damping (see Fig. \ref{fig:1}(a) on the structure of 3:2 resonances). This efficient damping helps in preventing spurious captures in unwanted high-order resonances.}
\label{fig:add:JoiningTheSystemTwoWays}
\end{figure}

\begin{table}[!h]
\centering
\begin{tabular}{|l|ll|}
\hline
\multicolumn{1}{|c|}{\multirow{2}{*}{$\Sigma_0 ~[\mathrm{g}/\mathrm{cm^2}]$}} & \multicolumn{2}{c|}{Success Probability}                                                                                                                                      \\ \cline{2-3} 
\multicolumn{1}{|c|}{}                                                        & \multicolumn{1}{c|}{\begin{tabular}[c]{@{}c@{}}8:5 -- 5:3\\ assembled\end{tabular}} & \multicolumn{1}{c|}{\begin{tabular}[c]{@{}c@{}}8:5 -- 5:3\\ not assembled\end{tabular}} \\ \hline
$5\times 10^{1}$	& \multicolumn{1}{l|}{0\%}	& 0\%	\\ \hline
$2.5\times 10^{2}$	& \multicolumn{1}{l|}{14\%}	& 22\%	\\ \hline
$5\times 10^{2}$	& \multicolumn{1}{l|}{22\%}	& 20\%	\\ \hline
$2.5\times 10^{3}$	& \multicolumn{1}{l|}{16\%}	& 4\%	\\ \hline
$5\times 10^{3}$	& \multicolumn{1}{l|}{22\%}	& 2\%	\\ \hline
\end{tabular}
\caption{}
\label{tbl:Prob}
\caption{{\bf Statistical study of the successful joining of the inner and outer system.}
We detail the success probabilities of joining the Trappist-1b,c,d Trappist-1,e,f,g subsystems for different surface densities $\Sigma_0$ (cfr.\ Eq.\ \eqref{eq:add:InnerDiscStruct}, with $r_0 = 0.025~ \mathrm{AU}$, ${\alpha_{\Sigma,0}}=3/5$). The left and right columns represent the extreme cases where the inner 8:5 -- 5:3 resonance between planets b, c, and d has already been established by the OLT on planet d or not, respectively.}
\label{tbl:Prob}
\end{table}

\clearpage

\setcounter{page}{1}
\cfoot{S \thepage}

\setcounter{figure}{0}
\renewcommand{\figurename}{Supplementary Figure}
\setcounter{table}{0}
\renewcommand{\tablename}{Supplementary Table}

\begin{sidewaystable}[!h]
\centering
\begin{tabular}{|l|l|l|l|l|l|l|}
\noalign{\vskip\doublerulesep
         \vskip-\arrayrulewidth} \cline{1-7}
Planet        & $m_{\mathrm{pl}}$ [$M_\oplus$]  & $R_{\mathrm{pl}}$ [$R_\oplus$]        & $T$ [days]                      & $e \cos\omega$ [$10^{-3}$] & $e \sin\omega$  [$10^{-3}$] & $i$ [${}^\circ$] \\ \hline\hline
b & $1.374\pm0.069$    & $1.116_{-0.012}^{+0.014}$  & $1.510826\pm6\times10^{-6}$     & $-2.15\pm3.32$ & $\hphantom{-}2.17\pm2.44$  & $89.728\pm0.165$   \\ \hline
c & $1.308\pm0.056$    & $1.097_{-0.012}^{+0.014}$  & $2.421937\pm1.8\times10^{-5}$   & $\hphantom{-}0.55\pm2.32$ & $\hphantom{-}0.01\pm1.71$  & $89.778\pm0.118$  \\ \hline
d & $0.388\pm0.012$    & $0.788_{-0.010}^{+0.011}$  & $4.049219\pm2.6\times10^{-5}$   & $-4.96\pm1.86$ & $\hphantom{-}2.67\pm1.12$  & $89.896\pm0.077$  \\ \hline
e & $0.692\pm0.022$    & $0.920_{-0.012}^{+0.013}$  & $6.101013\pm3.5\times10^{-5}$   & $\hphantom{-}4.33\pm1.49$ & $-4.61\pm0.87$ & $89.793\pm0.048$  \\ \hline
f & $1.039\pm0.031$    & $1.045_{-0.012}^{+0.013}$  & $9.207540\pm3.2\times10^{-5}$   & $-8.40\pm1.30$ & $-0.51\pm0.87$ & $89.740\pm0.019$  \\ \hline
g & $1.321\pm0.038$    & $1.129_{-0.013}^{+0.015}$  & $12.352446\pm5.4\times10^{-5}$  & $\hphantom{-}3.80\pm1.12$ & $\hphantom{-}1.28\pm0.70$  & $89.742\pm0.012$  \\ \hline
h & $0.326\pm0.020$    & $0.755_{-0.014}^{+0.014}$  & $18.772866\pm2.14\times10^{-4}$ & $-3.65\pm0.77$ & $-0.02\pm0.44$ & $89.805\pm0.013$  \\ \hline
\end{tabular}
\caption{Physical parameters of the Trappist-1 system (from \cite{2021PSJ.....2....1A}).}
\label{table:add:PlanetsInfo}
\end{sidewaystable}

\newpage

\begin{figure}[!h]
\centering
\includegraphics[width=1.\textwidth]{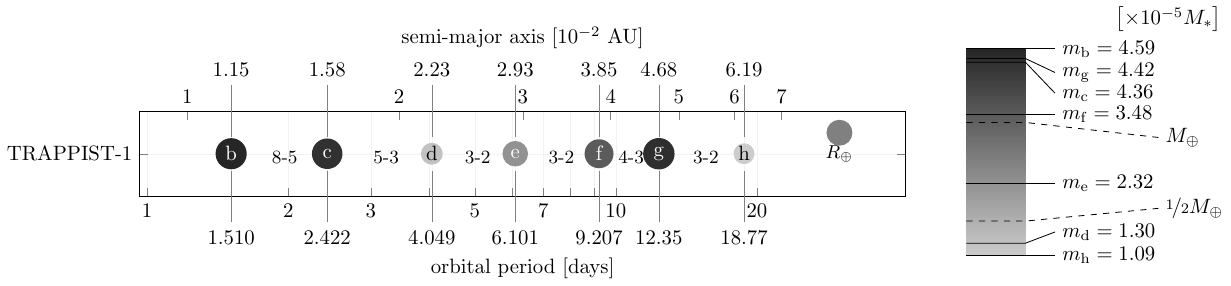}
\caption{Orrery of the Trappist-1 planetary system. The coloured circles represent each planet, with their size being a measure of the radii (the Earth radius is shown for reference in the upper-right corner). The shading of the circles represents the planets' masses, according to the legend on the right (half an Earth mass and 1 Earth mass are indicated for reference). The horizontal axes indicate the orbital period in days (bottom axis) and the semi-major axis in AU (top axis). The two-body resonances between the planets are also noted between the planets.}
\label{fig:add:PlanetsInfo}
\end{figure}
\begin{figure}[!tp]
\centering
\includegraphics[width=1.\textwidth]{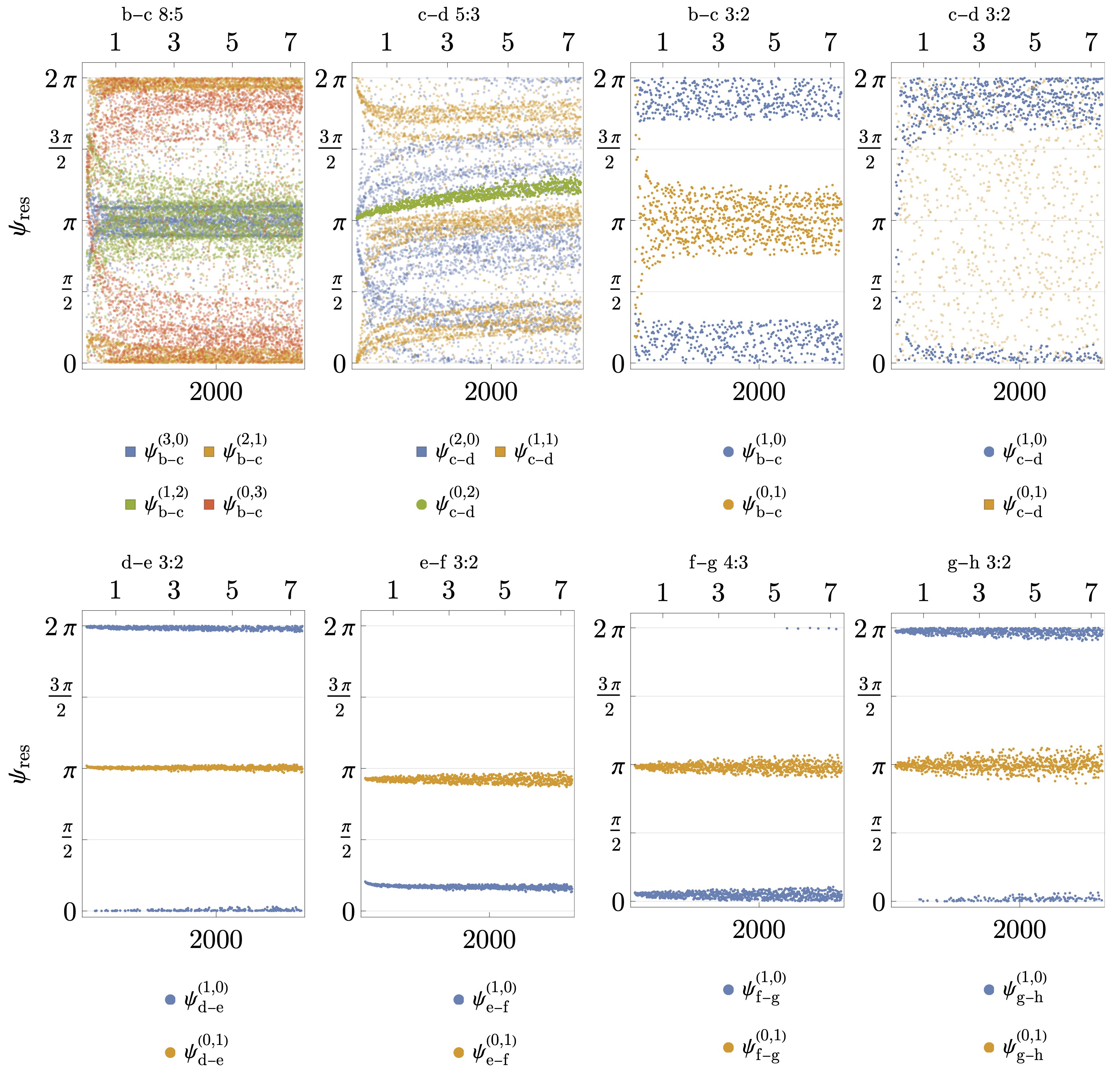}
\caption{All relevant two-body angles during the Gyr evolution of our simulated Trappist-1 system undergoing tidal dissipation. In each figure, the horizontal axes represent time in Gys (top axis) and time in circularisation timescales for planet b assuming tidal quality parameters $k_2/Q\sim 10^{-3}$ (bottom axis), as in Fig.\ \ref{fig:3}. The planet pair and the corresponding resonance is labeled above each plot, while the specific resonant angle is labeled below, using the notation introduced in Eq.\ \eqref{eq:ResonantAngles}. Slightly opaque markers indicate angles that show circulation at the end of the simulation after the effects of tidal dissipation (also indicated by coloured squares in the legends). For example, the 5:3 resonant angles $\psi_{\mathrm{c-d}}^{(2,0)} = 5{\lambda_{\mathrm{d}}}-3{\lambda_{\mathrm{c}}}-2{\varpi_{\mathrm{c}}}$ and $\psi_{\mathrm{c-d}}^{(1,1)} = 5{\lambda_{\mathrm{d}}}-3{\lambda_{\mathrm{c}}}-{\varpi_{\mathrm{c}}}-{\varpi_{\mathrm{d}}}$ are circulating, while $\psi_{\mathrm{c-d}}^{(0,2)} = 5{\lambda_{\mathrm{d}}}-3{\lambda_{\mathrm{c}}}-2{\varpi_{\mathrm{d}}}$ is librating until the end of the simulation. Libration of some 3:2 angles for pairs b-c and c-d is a geometrical effect due to the vanishing eccentricities $e_b$ and $e_c$ (the angles indeed circulate at the start of the simulation when the eccentricities have not been damped). The asymmetric libration of the 3:2 angles for the e-f system is due to the additional near-2:1 commensurability between e and g.}
\label{fig:add:2-body-angles}
\end{figure}

\begin{figure}[!tp]
\centering
\begin{subfigure}[b]{1. \textwidth}
\centering
\includegraphics[width=0.95\textwidth]{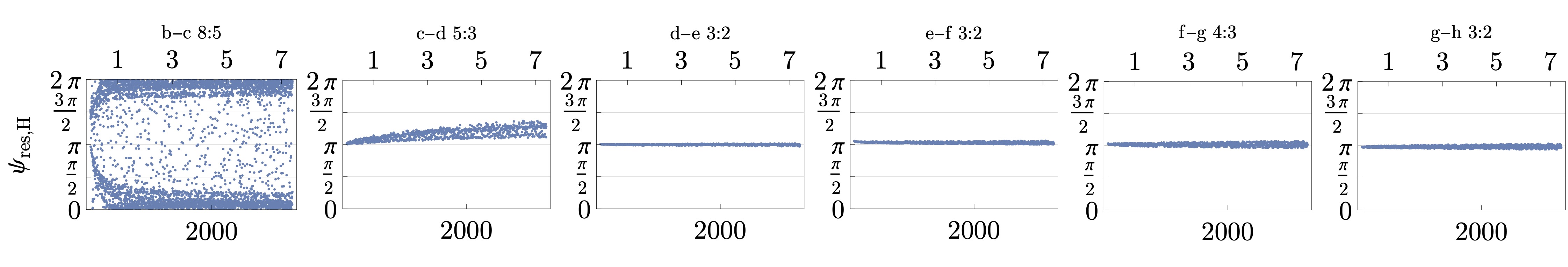}
\caption{}
\end{subfigure}
\begin{subfigure}[b]{1. \textwidth}
\centering
\includegraphics[width=.9\textwidth]{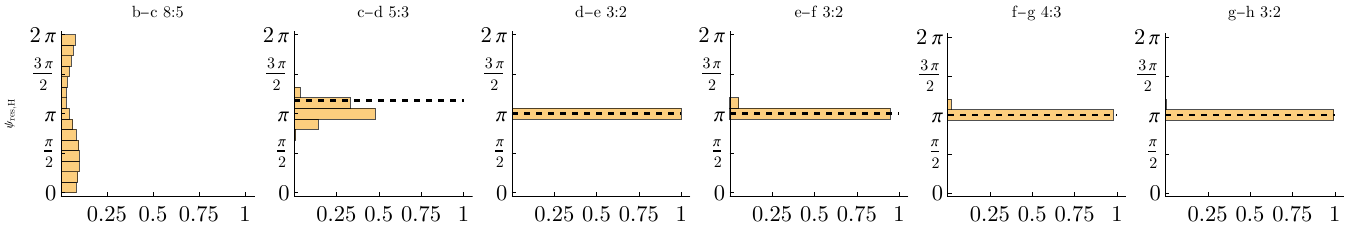}
\caption{}
\end{subfigure}%
\caption{Behaviour of the generalised two-body resonant angles introduced in \cite{2019AJ....158..238H}. Panel (a) shows their evolution taken from the same simulation as Fig.\ \ref{fig:3} and Supplementary Fig.\ \ref{fig:add:2-body-angles}. All these angles initially librate, and only the 8:5 angle for the Trappist-1b,c pair starts circulating as soon as $e_{\mathrm{c}}$ becomes vanishingly low. Panel (b) shows the distribution of the angles using the posterior from \cite{2021PSJ.....2....1A}. All angles except for the 8:5 angle are clustered near $\pi$. The dashed horizontal lines show the values of around which each angle librates at the end of our simulation from panel (a) for comparison (except in the case of the 8:5 angle for planets b and c, which circulates).}
\label{fig:add:HaddenAngles}
\end{figure}

\clearpage

{\bf Supplementary Discussion}\\

\ul{Physical properties of the Trappist-1 planets}\\
In our simulations we fix the masses and radii of the Trappist-1 planets to their mean values from \cite{2021PSJ.....2....1A}. Supplementary Table \ref{table:add:PlanetsInfo} lists the physical properties used, while Supplementary Fig.\ \ref{fig:add:PlanetsInfo} gives a graphical representation of the observed orbital architecture. As the mutual inclinations between the planets are negligible (less than $0.3^\circ$ including their uncertainties) we consider all planets on the same plane for simplicity. The Trappist-1 star has a mas of $0.0898\pm0.0023 M_\odot$ and a radius of $0.1192_{-0.0013}^{+0.0013} R_\odot$. In our nominal numerical experiments, we used the mean masses from \cite{2021PSJ.....2....1A}, but we also experimented with masses taken from a draw of the posterior distribution from \cite{2021PSJ.....2....1A}, and found similar results.\\

\ul{Calculation of one-sided Lindblad Torques}\\
A one-sided Lindblad torque (O-SLT) describes the exchange in angular momentum between a planet in a disc and the disc material that lies inside (inner Lindblad torque, ILT) or outside (outer Lindblad torque, OLT) the orbit of the planet. This exchange happens at specific locations in the disc, namely near Lindblad resonances, which are locations where 
\begin{equation}\label{eq:LindbladResonanceLocations}
   m (\Omega_{\mathrm{Kep}}(r) - \Omega_{{\mathrm{Kep}},{\mathrm{pl}}}) = \pm \kappa \sqrt{1+(m h)^2},
\end{equation}
with the plus sign in the case of the ILT and the minus sign for the OLT. Here, $\Omega_{\mathrm{Kep}}$ is the Keplerian frequency, $m$ is an integer indexing the Fourier component of the planet's gravitational potential, $\kappa$ is the epyciclic frequency, and $h=H/r$ is the disc's aspect ratio. 
Expressions for the one-sided Lindblad torques have been derived in multiple works \cite{1979ApJ...233..857G,1980ApJ...241..425G,1997Icar..126..261W,2008EAS....29..165M}. We note that they depend on the Laplace coefficients
\begin{equation}
    b_{1/2}^m(\alpha) : = \frac{2}{\pi} \int_0^{\pi} \frac{\cos (m \theta)}{(1-2\alpha \cos\theta +\alpha^2)^{1/2}} \D\theta
\end{equation}
and their derivatives $\frac{\D {}}{\D \alpha}b_{1/2}^m(\alpha)$, where $\alpha = r/r_{\mathrm{pl}}$. General closed-form formulas for the Laplace coefficients and their derivatives do not exist, and one typically resorts to numerical quadratures or approximation formulas \cite{2011MNRAS.418.1043Q,2020A&A...641A.176P}. 

Here, we note that the location where they need to be evaluated depends on the local aspect ratio $h$ of the disc (Eq.\ \eqref{eq:LindbladResonanceLocations}). In $N$-body integrations which include O-SLT, one would thus need to calculate the Laplace coefficients at each timestep, making it very inefficient. 
For a fixed aspect ratio across the disc, one can resort to tabulating the needed values \cite{2022MNRAS.511.3814H}, but for a more general implementation we have derived a concise approximation for the specific coefficients needed for the O-SLT calculation in the interest of efficiency and accuracy in the context of $N$-body simulations. 
To fix ideas it is customary to consider $0<\alpha<1$, since $b_{1/2}^m(\alpha)$ (and its derivative) when $\alpha>1$ can be expressed in terms of $1/\alpha<1$. We note that Laplace coefficients can be written in terms of hypergeometeric functions \cite{1995CeMDA..62..193L}
\begin{equation}
    b_{s}^m(\alpha) = 2 \frac{\Gamma(s+m) \alpha^m}{\Gamma(s)\Gamma(m+1)} {{}_{2}F_{1}}(s,s+m;m+1;\alpha^2);
\end{equation}
using the standard differentiation formula for ${{}_{2}F_{1}}$, one can also write for the derivative
\begin{align}
    \frac{\D {}}{\D \alpha}b_{1/2}^m(\alpha) &= 
        2m \frac{\Gamma(s+m) \alpha^{m-1}}{\Gamma(s)\Gamma(m+1)} {{}_{2}F_{1}}(s,s+m;m+1;\alpha^2) + \\
        &\quad\quad 2 \frac{\Gamma(s+m) \alpha^m}{\Gamma(s)\Gamma(m+1)} (2\alpha) \frac{s (s+m)}{m+1} {{}_{2}F_{1}}(s+1,s+m+1;m+2;\alpha^2).
\end{align}
For the case we are interested in, namely $s=1/2$, $\Gamma(1/2) = \sqrt{\pi}$ and the factor $\frac{\Gamma(1/2+m)}{\Gamma(m+1)}$ can be tabulated for the needed number of indices and thus quickly accessed in $N$-body integrations. For the term ${{}_{2}F_{1}}(1/2,1/2+m;m+1;\alpha^2)$, using the Pfaff transformation ${{}_{2}F_{1}}(a,b;c;z) = (1-z)^{-a} {{}_{2}F_{1}}\left(a,c-b;c;\frac{z}{z-1}\right)$ we write
\begin{equation}
{{}_{2}F_{1}}(1/2,1/2+m;m+1;\alpha^2) \equiv (1-\alpha^2)^{-1/2} {{}_{2}F_{1}}\left(1/2,1/2;m+1;\frac{\alpha^2}{\alpha^2-1}\right),
\end{equation}
and the factor still depending on the hypergeometric function can be expanded in series:
\begin{equation}
{{}_{2}F_{1}}\left(1/2,1/2;m+1;\frac{\alpha^2}{\alpha^2-1}\right) = 1 + \frac{\left(\frac{1}{2}\right)\left(\frac{1}{2}\right)}{m+1} \frac{\left(\frac{\alpha^2}{\alpha^2-1}\right)}{1!} + ...;
\end{equation}
similar calculations apply for the ${{}_{2}F_{1}}(3/2,3/2+m;m+2;\alpha^2)$ term that appears in the derivative of the Laplace coefficients.
Additional terms in the expansions can be added at a relatively small computational cost, but we found that the first two terms are sufficient, making an approximate but explicit expression of the Laplace coefficients and their derivatives extremely concise. We note that, for a fixed $m$, this approximation gets worse and worse as $\alpha\to 1$, but the approximation remains good up until $\alpha$ values closer to 1 for larger fixed $m$. This is favourable, since for O-SLT an $\alpha\to1$ corresponds to higher $m$ values, and the limit $\alpha\to1$ is actually never fully reached as the Lindblad resonant locations \eqref{eq:LindbladResonanceLocations} cannot accumulate arbitrarily close to the location of the planet due to pressure support. Indeed, we checked that, when evaluated at the Lindblad resonance locations for reasonable aspect ratios $h$, the relative error of this approximation to the real Laplace coefficients is always less than $\sim 5\%$, and that of their derivatives is always less than $\sim 10\%$.

Isolated O-SLT associated with indices $m$ must be added up to get the full torque: in the case of the outer Lindblad torque one has
\begin{equation}\label{eq:FullOLT}
    \Gamma_{{\mathrm{OLT}}} = \sum_{m=1}^{m_{\mathrm{max}}} \Gamma_{{\mathrm{OLT}}, m},~ m_{\mathrm{max}}\sim 1/h.
\end{equation}
We note that, unlike \cite{2022MNRAS.511.3814H}, we typically need to consider $m$'s larger than 1 (the outer 2:1 Lindblad resonance) since in our simulations the planets find themselves very close to the inner edge of the disc. In the case of a planet inside the inner cavity, where the surface density drops at smaller separations from the star, the individual torques' strengths decrease with $m$, as they are proportional to the local surface density. One can therefore stop the summation when the additional contribution does not constitute a major refinement to the already accumulated torque from smaller $m$'s: we choose a threshold of a 10\% difference to determine this cutoff, as this is the error in our Laplace coefficients calculation anyway; besides, planet-disc interaction formulas widely used in population synthesis have errors up to 20\% \cite{2011MNRAS.410..293P,2023A&A...670A.148P}. For a planet close to the inner edge, we typically end up summing around $m_{\mathrm{max}} = 15$ terms, which is where the absolute value of the individual O-SLT's start decaying \cite{1997Icar..126..261W,2008EAS....29..165M}, and is close to $m_{\mathrm{max}}\sim 1/h$ \cite{2008EAS....29..165M} for reasonable disc aspect ratios. 

Finally, one must model the transition between the full torque felt by a planet when it is completely inside the disc and the outer Lindblad torque it experiences when it is slightly inside the cavity. For simplicity, we chose a smooth function such as \eqref{eq:tildeG} to transition between the two regimes.
\\

\ul{Evolution of the Trappist-1b,c,d,e inner system: capture in the 3:2 chain and clearing of the inner disc}

In our model, we consider an inner system composed by planets Trappist-1b, c, d and e that is initially dynamically separate from the outer system (the idea of a separation of an inner and outer Trappist-1 system has already been investigated in \cite{2018MNRAS.476.5032P,2022MNRAS.511.3814H}). \cite{2022A&A...658A.170T} showed that, under customary disc-driven type-I migration, these planets would naturally assemble in a 3:2 resonance chain at the inner edge of the disc. In this section, we give an example of the assembly of such resonant chain, followed by a phase in which the inner disc clears to leave planets b, c and d inside an inner cavity, while planet e remains inside the disc. This step serves the purpose to generate initial conditions for the $N$-body integrations shown in Fig.\ \ref{fig:1} (panels b, c, d).

As initial conditions, we assume that the planets are wide of the 3:2 -- 3:2 -- 3:2 chain, but already inside the 2:1, since the 3:2 is the most likely chain formed by the planets \cite{2022A&A...658A.170T}, and with initially vanishing eccentricities because of the damping provided by the disc. The planets thus find themselves at the right-bottom corner of Fig.\ \ref{fig:1}(a), and the system undergoes NAM decrease because of the negative torque provided by the disc. They thus smoothly follow the equilibrium curves until the $e$-damping provided by the disc counterbalances the push provided by the torque along the resonant curves \cite{2018CeMDA.130...54P}. With this evolution, capture in a 3:2 -- 3:2 -- 3:2 resonance chain is then ensured by adiabatic theory. 

Various mechanisms can cause the planets to fall inside the cavity, such as the opening of a partial gap for thin/low-viscosity discs, or an enlargement of the magnetic cavity due to activity of the star. More specifically, if planets b and c, being rather massive, can open large gaps, planet e, being more massive than f, can be able to push it inside the disc inner edge \cite{2021A&A...648A..69A}. One way to reproduce the scenario in which planets b, c and d find themselves inside the inner edge of the disc using an $N$-body simulations is to have their surrounding gas removed.
We assume that the clearing of the disc happens on a timescale that allows for an adiabatic evolution of the chain, that is, the timescale of resonant evolution is shorter than the timescale of disc evolution at this stage. This requirement is actually necessary in any model for the formation of any resonant chain: the removal of the disc cannot be too fast as to disrupt the system. In practice, however, this is not a restrictive assumption because the physical evolution of the disc and star are slow compared to the planets' relevant dynamical timescales. We also note that, if planet e sits at the inner edge while planets b, c and d are in the cavity with planet d feeling an OLT, it would take a time of order $10^{7}$ years to for these planets to reach the 8:5 -- 5:3 period ratio from an initial 3:2 resonant chain. This is a considerable longer time than in the model of \cite{2022MNRAS.511.3814H} (where instead planets b and c are in the cavity and planet d sits at the inner edge) since the OLT on planet d (mostly the sole $m=1$ component) would alone need to push all three planets inside the cavity (we recall that the one-sided Lindblad torque is proportional to the planet's mass squared, as well as the disc surface density and $h^{-3}$, \cite{2008EAS....29..165M}). Thus, in our model, the time over which planets b, c and d fall inside the inner cavity cannot be arbitrarily long. However, the clearing of a gap around planets and/or the pushing of planets inside the inner edge are observed to happen within timescales of a few $10^{3} - 10^{5}$ orbits in hydrodynamical simulations (e.g.\ \cite{2006Icar..181..587C,2021A&A...648A..69A,2023A&A...670A.148P}).

After the resonance is established, we model this phase in our simulation by removing the gas in the inner portion of the disc as described in Eqs.\ \eqref{eq:add:InnerDiscRemoval1},\eqref{eq:add:InnerDiscRemoval2}. We choose $r_{{\mathrm{ed}},{\mathrm{in}}}$ as the original inner edge $r_{{\mathrm{ed}}}$, while $r_{{\mathrm{ed}},{\mathrm{fin}}}$ is taken to be at a location between planet d and e, in our nominal scenario. At this point, planets b to d are in a disc cavity, subjected to tidal evolution (and OLT for the outermost planet) while planet e is still in the disc and is subject to its unchanged (negative) torque, which drives it at the location of the inner edge of the disc.
We stress that, as long as the disc evolution is adiabatic, the specific way in which the disc clears out is irrelevant, and the system will keep following a resonant locus like the ones plotted in Fig.\ \ref{fig:1}.\\
Extended Data Fig.\ \ref{fig:add:bcdeBuild} reproduces the full evolution of the Trappist-1b,c,d,e system from the capture into the resonance until the entrance of planets b, c and d into the inner cavity, obtained via the removal of the inner part of the disc.
This final state is used as initial condition for the simulation depicted in Fig.\ \ref{fig:1}. In passing, we note that the information on the initial eccentricities $e_{\mathrm{capt}}\propto h$ (where $h$ is the aspect ratio of the disc) that the planets acquired at the moment of the capture in the 3:2 -- 3:2 -- 3:2 resonant chain (i.e.,\ the primordial NAM, which depends on the migration and disc conditions) is lost once planet e starts receding with the inner edge, and we are thus not sensitive to different initial disc structures. This is because, if for instance the eccentricities had been higher, the system's evolution under NAM increase must follow the same tracks shown in Fig.\ \ref{fig:1}, so they would reach the lower eccentricity states and the same dynamical state, and when planets b-c-d reach the 8:5 -- 5:3 period ratio the system is in the exact same orbital state regardless of the initial capture eccentricities.\\

\ul{Joining the inner and the outer systems}

In this section we detail slightly different dynamical pathways in the assembly of the full Trappist-1 chain.
Our preferred scenario is that planets f and g approached e while having already formed their 4:3 resonance. The combined push of planets f-g to planet e due to the e-f 3:2 resonance is enough to dislodge e from the inner edge of the disc. This is because 1) planet e's eccentricity would be large, $e_{\mathrm{e}}\simeq 0.04\sim h_0$ (while it only reaches 0.003 when b-c-d cross the 8:5 -- 5:3 resonance), since it is the least massive body in the e-f-g sub-system; 2) the positive corotation torque is quenched when a planet's orbit is excited \cite{2014MNRAS.437...96F,2021A&A...648A..69A}; and 3) the one sided Lindblad torque is larger than the total Lindblad torque by a factor comparable to the scale height of the disc \cite{2008EAS....29..165M}, and the mass ratios between planets e, f and g are so that planet e would migrate inward due to the OLT faster than planets f and g. We note that this cannot hold true indefinitely, because, as planet e migrates inward, the number of individual OLT that remain active (i.e.,\ the number of non-vanishing addends in Eq.\ \eqref{eq:FullOLT}) decreases, and the OLT looses steam. Instead, the torque on the outer planets still inside the disc remains unaffected. For this reason, planet f will always catch up with planet e and eventually re-establish a 3:2 resonance with it.
We also note that our understanding of the strength of the Lindblad and corotation torques at the inner edge of the disc is limited. In particular, the corotation torque may also become one-sided \cite{2017A&A...601A..15L}. The exact details of these regimes of planet-disc interaction are outside the scope of this work; we remark, however, that \cite{2021A&A...648A..69A} show with hydrodynamical simulations that planets with planet mass to stellar mass ratios of the order of the Trappist-1 system and interacting in mean motion resonances observed in the Trappist-1 chain can indeed be pushed inside the disc inner cavity.

The joining of the inner and outer systems is intrinsically a probabilistic process due to the nature of high order resonances. 
Two extremes are possible: the inner Trappist-1b,c,d system has had time to assemble a full 8:5 -- 5:3 chain (via OLT-driven convergent migration, where the OLT is active on planet d) or not. This will depend on a combination of the time elapsed between the divergent crossing of the 8:5 -- 5:3 commensurability (Fig.\ \ref{fig:1}) and the arrival of planet e (Extended Data Fig.\ \ref{fig:add:JoiningTheSystemTwoWays}), and the strength of the OLT (i.e.,\ the surface density at the inner edge). These quantities are not constrained, so we consider both extremes to evaluate the probability of the outcomes.
We show in Extended Data Table \ref{tbl:Prob} the probability of a successful assembly of the observed Trappist-1 chain under different disc surface densities $\Sigma_0$ at the location of the inner edge, in both these extremes. For each setup, this probability is calculated by running 50 simulations with different initial conditions where the phase of the angle at the approach of planet e to planet d is varied uniformly. If the surface density is of order $10^{2}~ \mathrm{g}/\mathrm{cm^2}$ or higher, the assembly of a Trappist-1 chain is always possible. 
The probability of successfully assembling a Trappist-1 chain in the case where the 8:5 -- 5:3 chain between planets b, c and d has not been fully assembled is similar to the case where it has been fully assembled for surface densities $\Sigma_0$ of order $10^{2} ~\mathrm{g}/\mathrm{cm^2}$ at the location of the inner edge, reduced for $\Sigma_0$ of order $10^{3} ~\mathrm{g}/\mathrm{cm^2}$, but this is unlikely for higher surface densities, due to the more delicate nature of the b-c-d subsystem. Since higher surface densities at the inner edge will speed up the assembly of a true 8:5 -- 5:3 inner resonance, under the assumptions of our model we expect that, in the real system, the probability of assembling the currently observed resonance was in between the values shown in the two columns in Extended Data Table \ref{tbl:Prob} for the higher surface densities, while it would have been closer to the ones shown in the right column of Extended Data Table \ref{tbl:Prob} for the lower surface densities.
{For a surface density with highest probability of success, $\Sigma_0 = 5\times 10^{2}~ \mathrm{g}/\mathrm{cm^2}$, we also tested an aspect ratio of $h_0=0.04$ (a colder disc), and the probability was 12\%. 
This is due mainly to the stability of the inner b-c-d system, and can be explained as follows. The $m = 1$ component of the OLT (felt by planet d) is relatively similar for the two aspect ratios $h$, while the full OLT (felt by planet e while approaching planet d) is stronger for smaller $h$ (since it scales with $h^{-3}$, e.g.\ \cite{2008EAS....29..165M}); this means that the relative migration speeds between planets d and e are different in the two cases in such a way that planet e will approach planet d slightly faster for lower $h$, which reduces the stability of the inner 8:5 - 5:3 chain.
Similarly to the question of the one-sided corotation torque mentioned above, we note that, despite our dedicated efforts to develop a realistic description in the Method section \emph{Calculation of one-sided Lindblad Torques}, our prescription of the OLT for planets inside the inner edge of the disc may be subject to improvements in the light of future work such as tailored hydrodynamical simulations.}

Different approach histories are still possible. We tested scenarios where planets f and g approach planet e while deep in their 4:3 resonance, or slightly outside it so that f captures e in the 3:2 before a 4:3 between f and g is established, and both yield similar outcomes (see Extended Data Fig.\ \ref{fig:add:JoiningTheSystemTwoWays}).

After the joining of the inner b-c-d-e system with the f-g system, the last planet h joins the resonant chain at some later time (displayed as a cut in the time axis in Fig.\ \ref{fig:2}). Planet h most likely reached the inner regions after planets f and g since it is less massive and thus it migrates slower than planets f and g. We again assume that this does not happen after the inner edge has moved out excessively. \cite{2022A&A...658A.170T} showed that planet h would likely capture in a 2:1 resonance with planet g under realistic migration prescriptions; based on the observed 3:2 resonance between these planets, \cite{2022MNRAS.511.3814H} argue for a formation imprint regarding planets g and h, in that planet h forms inside the 2:1 resonance with g in order to ensure the capture in the 3:2 resonance.

Another possibility is that, after f reaches the inner edge of the disc, it gets pushed inside the cavity by g, and the inner edge position is then obtained by g. This is however not likely because the mass difference between planet f and g is not as large as the one from planet e and the combined f-g system. Still, we checked that this scenario can result in a successful assembly of the correct resonant chain, with the difference that the final eccentricities are extremely low, up to $10^{-3}$, even when the gas is still present. This is due to the fact that, in this scenario, only planet h experiences an inward push due to its (full) torque: however, since planet h is the least massive, the push it can provide is much lower than that of g, so that the final resonant state is extremely relaxed. A resonant state with such low eccentricities already at the end of the gas phase, comparable with the currently observed ones (Fig.\ \ref{fig:3}), is hard to reconcile with the subsequent evolution after disc dispersal, unless virtually no tidal damping was provided over many Gyrs.\\

\ul{Planets' formation history and physical characterisation}

The Ansatz that planets f, g and h migrate to the inner part of the disc at a later time naturally involves a condition on the timescale over which this takes place. In our model, this is realised in a simple relation between the (average) rate at which the inner edge moves outward after the crossing of the 8:3 -- 5:3 b-c-d resonance carrying planet e with it, which can be arbitrarily slow, and the rate at which the planets in the outer sub-system form and migrate inward. More specifically, after the 8:3 -- 5:3 resonance crossing, the inner edge position $r_\mathrm{ed}$ should not have moved outwards by more than $\sim 7\%$ before the arrival of planet f, to ensure that the period ratio between e (which is carried away by the inner edge) and d (whose position is relatively fixed, inside the disc cavity) does not exceed 2. If $t_{\mathrm{form},\mathrm{f}}$ is the time it takes for planet f to have formed and migrated to the inner edge of the disc after the time $t_\mathrm{cross}$ at which planets b-c-d divergently crossed the 8:3 -- 5:3 b-c-d resonance, then the (average) inner edge recession timescale $\tau=r_{{\mathrm{ed}}}(t_\mathrm{cross})/(\mathrm{d}r_{{\mathrm{ed}}}/\mathrm{d} t)$ should be longer than $\sim10 t_{\mathrm{form},\mathrm{f}}$.
Based on the model of \cite{2017A&A...604A...1O}, assuming that protoplanets form and grow one by one at the iceline, \cite{2021ApJ...907...81L} found that the formation and growth timescale may be linked to timing of embryo formation. The inference that the inner planets are water poor \cite{2021PSJ.....2....1A,2022NatAs...6...80R} could then be explained by post-formation processes \cite{2017MNRAS.464.3728B,2020ApJ...889...77H}.
Another possibility is that planets b-c-d-e and f-g-h formed at two distinct locations in the disc, one where rocky planetesimals and one where icy planetesimals formed, respectively \cite{2022NatAs...6...72M,2023NatAs...7..330B}. This could explain how the outer planets joined the inner system at a later time, and at the same time would have consequences for the compositions acquired during the formation stage. In particular, the two-location origin would be consistent with the inference that planets f-g-h are rich in water, while b-c-d are rocky (\cite{2021PSJ.....2....1A,2022NatAs...6...80R}, as well as direct observations for planets b and c, \cite{2023Natur.618...39G,2023Natur.620..746Z}). However, planet e may also be water rich. This would not necessarily be inconsistent with our scenario. Planet e could have started to form rocky, but, remaining longer at the disc’s inner edge than planets b, c and d, it may have accreted water-rich vapour from the disc \cite{2021A&A...649L...5B}; it may also have acquired enough water from late accretion after the dispersal of the system’s gaseous disc \cite{2022NatAs...6...80R}, or water may be produced through the oxidation of $\mathrm{H}_2$ in the planetary atmosphere by oxides available at the surface \cite{2006ApJ...648..696I}. Our understanding of the composition of the Trappist-1 planets remains limited, and future characterisation will give further insights into the system's formation history.
\\

\ul{Tidal evolution of a full 7-planet chain} \\
Previous simulations of the full 7-planet Trappist-1 system have shown that the preferred outcome of planet-disc interactions is the formation of a 3:2 -- 3:2 -- 3:2 -- 3:2 -- 4:3 -- 3:2 chain \cite{2022A&A...658A.170T}. After the dispersal of the disc, one might attempt to recover the current state of the system via tidal evolution alone. However, this mechanism alone is not sufficient to recover the full Trappist-1 chain \cite{2022MNRAS.515.2373B,2022MNRAS.511.3814H}. Indeed, the orbits of neighbouring planets may expand relative to each other only if all orbits along the chain are subjected to the same fate (see Methods). 
Thus, if the inner three planets had tidally evolved from a 3:2 -- 3:2 configuration into a significantly wider one (closer to the observed 8:5 -- 5:3), so would the outer four planets. Instead, the outer four planets reside extremely close to their respective first order resonances, ruling out this possibility.
This shows that starting from the preferred 3:2 -- 3:2 -- 3:2 -- 3:2 -- 4:3 -- 3:2 chain and subsequently evolving the system once the disc has fully dissipated cannot explain the currently observed orbital configuration.\\

\ul{Dissipative evolution of the Trappist-1b,c,d,e inner system: resonant repulsion}\\
In this section we describe the outcome of a simulation in which the inner Trappist-1b,c,d,e system starts from the nominal 3:2 -- 3:2 -- 3:2 resonance and where the resonant repulsion is driven by dissipative (e.g.\ tidal) $e$-damping. Extended Fig.\ \ref{fig:tide_repuls} shows the evolution of the period ratios and the eccentricities over time (in units of circularisation timescales of planet b), for whatever choice of tidal parameters. As expected, the evolution is exactly the same as the one shown in Fig.\ \ref{fig:1}, up until planets b-c-d simultaneously cross the 8:5 -- 5:3 resonance. At this point, the eccentricities are temporarily excited, but the $e$-damping provided by the dissipative force re-establishes the 3:2 resonances between planets b, c and d, and the system continues to evolve away from the 8:5 -- 5:3 -- 3:2 resonances. Namely planets b, c and d recover their resonant repulsion (their period ratios continue to increase) while planet e is now not pushed away by the 3:2 resonance anymore and does not evolve in semi-major axis, so that $T_{\mathrm{e}}/T_{\mathrm{d}}$ decreases ($T_{\mathrm{d}}$ increases as planet d is being pushed outwards by the 3:2 resonance with planets c and b).
This shows that an outward moving inner edge is a more robust mechanism to drive resonant repulsion at this stage.\\

\ul{Description of the final resonant state}\\
We give here a description of the resonant state observed at the end of our simulations and how it relates to the observed state.

Supplementary Fig.\ \ref{fig:add:2-body-angles} shows the evolution of the two-body resonant angles across the Trappist-1 chain. Keeping in mind that the observed period ratios are extremely close to the 8:5 and 5:3 ratios, whether or not an angle will be librating depends on the eccentricities of the planets. This is because at the end of the simulation $e_{\mathrm{b}}$ and $e_{\mathrm{c}}$ are low (consistent with vanishing eccentricity) so ${\varpi_{\mathrm{b}}}$ and ${\varpi_{\mathrm{c}}}$ are fast; instead $e_{\mathrm{d}}$ is higher, so ${\varpi_{\mathrm{d}}}$ is slow. This means that the 5:3 between c and d is split in three locations in the period ratio space (given by the libration condition for the three resonant angles $\psi_{\mathrm{c-d}}^{(2,0)} = 5{\lambda_{\mathrm{d}}}-3{\lambda_{\mathrm{c}}}-2{\varpi_{\mathrm{c}}}$, $\psi_{\mathrm{c-d}}^{(1,1)} = 5{\lambda_{\mathrm{d}}}-3{\lambda_{\mathrm{c}}}-{\varpi_{\mathrm{c}}}-{\varpi_{\mathrm{d}}}$, $\psi_{\mathrm{c-d}}^{(0,2)} = 5{\lambda_{\mathrm{d}}}-3{\lambda_{\mathrm{c}}}-2{\varpi_{\mathrm{d}}}$; see Methods section). Only the angle $\psi_{\mathrm{c-d}}^{(0,2)} = 5{\lambda_{\mathrm{d}}}-3{\lambda_{\mathrm{c}}}-2{\varpi_{\mathrm{d}}}$ (shown in green in the second panel of Supplementary Fig.\ \ref{fig:add:2-body-angles}) has a slow ${\varpi_{\mathrm{d}}}$ and the resonance is located close to a 5:3 period ratio.
Instead, for the 8:5 between b and c, all resonant centres are displaced, so no resonant angle can librate close to the 8:5 period ratio, and the angles indeed show sporadic but periodic circulations at the end of the simulation. We note however that, before the effects of tidal dissipation manifest themselves, $e_{\mathrm{c}}$ is initially non-vanishing (Fig.\ \ref{fig:3}(a)). Thus, all 5:3 resonant angles for the c-d pair and the angle $\psi_{\mathrm{b-c}}^{(0,3)} = 8{\lambda_{\mathrm{d}}}-5{\lambda_{\mathrm{c}}}-3{\varpi_{\mathrm{c}}}$ (in red on the top-left panel of Supplementary Fig.\ \ref{fig:add:2-body-angles}) are librating at the beginning of the simulation. This shows that the currently observed 8:5 -- 5:3 ratios for planets b, c and d are indeed a result of a past true evolution in these resonances which was established during the disc-phase assembly of the system, and the subsequent spurious circulation of some of these angles is just a result of the tidal evolution.

We also note that some of the 3:2 angles for b-c and d-c are observed to be librating. This is a geometrical effect due again to the low eccentricities for planets b and c \cite{2012A&A...546A..71D}, and indeed the angles are initially circulating when the eccentricities are large. Instead, since $e_{\mathrm{d}}$ remains large, the angle $\psi_{\mathrm{c-d}}^{(0,1)} = 3{\lambda_{\mathrm{d}}}-2{\lambda_{\mathrm{c}}}-{\varpi_{\mathrm{d}}}$ is observed to be circulating (shown in orange in the fourth panel of Supplementary Fig.\ \ref{fig:add:2-body-angles}).
This is indeed a different state than the initial 3:2 -- 3:2 chain captured via type-I migration. Here capture is aided/driven by tidal damping, which is asymmetric: $1/\tau_{e,\mathrm{tides};{\mathrm{b}}}$, $1/\tau_{e,\mathrm{tides};{\mathrm{c}}}$ are higher (more efficient tides), while $1/\tau_{e,\mathrm{tides};{\mathrm{d}}}$ is lower (more inefficient tides). After the kick in $e$ at the simultaneous 8:5 -- 5:3 crossing, the movement is mostly vertical in a period ratio vs.\ eccentricity plane (Fig.\ \ref{fig:1}), so the only angle that can start librating is the 5:3 angle $5{\lambda_{\mathrm{d}}}-3{\lambda_{\mathrm{c}}}-2{\varpi_{\mathrm{d}}}$.

2-body resonant angles at low eccentricities typically only have symmetric librations, meaning that they librate around the values 0 and $\pi$ \cite{2018CeMDA.130...54P}, as we see in most angles in Supplementary Fig.\ \ref{fig:add:2-body-angles}. A notable exception is the libration of the e-f 3:2 resonant angles, specifically $\psi_{\mathrm{e-f}}^{(1,0)} = 3{\lambda_{\mathrm{f}}}-2{\lambda_{\mathrm{e}}}-{\varpi_{\mathrm{e}}}\simeq \pi + \theta$, $\theta\simeq -0.8\pi$. This is due to the additional near-resonance between planet e and g, which are close to the 2:1 ratio by virtue of the e-f-g system being in a 3:2 -- 4:3 chain (a similar effect is observed in Kepler-223 which contains the same sub-resonant chain, \cite{2017A&A...605A..96D,2021AJ....161..290S}). These asymmetric libration islands always come in pairs because the resonant terms only contain cosines of the angles by the d'Alembert rules, and thus the phase space is perfectly symmetric around $\psi=\{0,\pi\}$ axis. For this reason, libration around $\pi+\theta$ and $\pi-\theta$ is dynamically equivalent and capture into either resonant island is a stochastically equivalent event. \\

Although the libration of the two-body angles rather than the period ratio alone is a confirmation of a resonant state, the former are hard to observe due to uncertainties in the longitudes of pericentres of the planets. The generalised resonant argument introduced by \cite{2019AJ....158..238H} are less sensitive to these uncertainties. 
Panel (a) of Supplementary Fig.\ \ref{fig:add:HaddenAngles} shows the evolution of the generalised resonant argument for each planet pair for the full 8:5 -- 5:3 -- 3:2 -- 3:2 -- 4:3 -- 3:2 chain from the same simulation as in Fig.\ \ref{fig:3} and Supplementary Fig.\ \ref{fig:add:2-body-angles}. All angles are initially librating, but the angle associated with the 8:5 resonance between b and c quickly begins circulating as soon as $e_{\mathrm{c}}$ becomes vanishingly low. Panel (b) shows their distribution taken from the posterior from \cite{2021PSJ.....2....1A}, showing a clustering of these angles around the predicted values from our simulation, except for the 8:5 angle, which attains values in the whole interval $(0,2\pi)$ like in our simulated system.\\

\ul{A Trappist-1b,c,d inner system}\\
Repeating the same exercise of drawing equilibrium curves for a 3-planet 3:2 resonance inner chain, one finds that many of the same arguments apply (see also \cite{2022MNRAS.511.3814H,2018MNRAS.476.5032P} in the context of their models). Specifically, it remains true that, along the curve of resonant equilibria which link all period ratios and eccentricities to each other, the pair b-c crosses the 8:5 resonance at the same time as the pair c-d crosses the 5:3. Thus the system could in principle be split into an inner Trappist-1b,c,d, divergently evolving from a 3:2 -- 3:2 configuration into a 8:5 -- 5:3 along the resonant curve, and outer Trappist-1e,f,g,h system in a 3:2 -- 4:3 -- 3:2 chain. If the resonant repulsion mechanism in the inner system is driven by tides we are again left with having to resort to fine-tuning of timescales and tidal parameters, and similarly if the repulsion is driven by a OLT acting onto c (see main text).
We can instead again resort to the idea of a receding inner edge, where now planet d resides. However, as the inner edge would otherwise continue to recede and there is no reason to believe that it would stop doing so exactly when the 8:5 -- 5:3 resonance is crossed, one must invoke a mechanism to dislodge planet d from the inner edge in order to lock c and d at the 5:3 resonance. Although this is not inconceivable (for example the reduction of the corotation torque due to an increase in eccentricity), planet d would now find itself right next to the inner edge of the disc and thus experiences the full OLT, instead of just the $m=1$ Fourier component, like in the case when e was at the disc inner edge. Planet d's inward migration would then be much stronger than if the inner edge was farther away, and so planet d might well skip the 5:3 resonance with c and migrate further in. Although possible, considering the same picture with only the inner three planets introduces a limitation for parameters that allow for the formation of the Trappist-1 chain. Namely, two new requirements are needed: i) that the OLT win over the reduced corotation torque at the 5:3 resonance crossing due to the kick in $e$; and ii) that the combined OLT felt by planet d be small enough to allow for an immediate capture into the 5:3. These requirements are entirely absent in the case of an inner system including planets b to e.
The only advantage of considering the inner system as comprised of planets b, c and d only is that the compositional dichotomy seems more straighforward to explain. Indeed planets b to d are thought to be water poor, while planets e to f to be water rich \cite{2022NatAs...6...80R}. However, in our preferred scenario with planet e at the inner edge, planet e has been surrounded by the gaseous environment for longer than planets b to d, so that it could have accreted later icy pebbles or a larger envelope. Indeed, its mass is twice as large as planet d. Planets b and c large masses may instead be the result of collisions \cite{2022A&A...658A.184O}.\\

\renewcommand\refname{Supplementary Information References}


\begin{thebibliography}{9}

\bibitem
{2017NatAs...1E.129L} Luger, R. et al.\ 2017. A seven-planet resonant chain in TRAPPIST-1. \nastro\ 1. doi:10.1038/s41550-017-0129

\bibitem
{2021PSJ.....2....1A} Agol, E. et al.\ 2021. Refining the Transit-timing and Photometric Analysis of TRAPPIST-1: Masses, Radii, Densities, Dynamics, and Ephemerides. \psj\ 2. doi:10.3847/PSJ/abd022

\bibitem
{2012ARA&A..50..211K} Kley, W., Nelson, R.~P. 2012. Planet-Disk Interaction and Orbital Evolution. \araa\ 50, 211–249. doi:10.1146/annurev-astro-081811-125523

\bibitem
{2017ApJ...840L..19T} Tamayo, D., Rein, H., Petrovich, C., Murray, N. 2017. Convergent Migration Renders TRAPPIST-1 Long-lived. \apj\ 840. doi:10.3847/2041-8213/aa70ea

\bibitem
{2022A&A...658A.170T} Teyssandier, J., Libert, A.-S., Agol, E. 2022. TRAPPIST-1: Dynamical analysis of the transit-timing variations and origin of the resonant chain. \aap\ 658. doi:10.1051/0004-6361/202142377

\bibitem
{2022MNRAS.515.2373B} Brasser, R., Pichierri, G., Dobos, V., Barr, A.~C.\ 2022.\ Long-term tidal evolution of the TRAPPIST-1 system.\ \mnras\ 515, 2373–2385. doi:10.1093/mnras/stac1907

\bibitem
{2018MNRAS.476.5032P} Papaloizou, J.~C.~B., Szuszkiewicz, E., Terquem, C. 2018. The TRAPPIST-1 system: orbital evolution, tidal dissipation, formation and habitability. \mnras\ 476, 5032–5056. doi:10.1093/mnras/stx2980

\bibitem
{2022MNRAS.511.3814H} Huang, S., Ormel, C.~W. 2022. The dynamics of the TRAPPIST-1 system in the context of its formation. \mnras\ 511, 3814–3831. doi:10.1093/mnras/stac288

\bibitem
{2012ApJ...756L..11L} Lithwick, Y., Wu, Y.\ 2012.\ Resonant Repulsion of Kepler Planet Pairs.\ \apj\ 756. doi:10.1088/2041-8205/756/1/L11

\bibitem
{2013AJ....145....1B} Batygin, K., Morbidelli, A.\ 2013.\ Dissipative Divergence of Resonant Orbits.\ \aj\ 145. doi:10.1088/0004-6256/145/1/1

\bibitem
{2019A&A...625A...7P} Pichierri, G., Batygin, K., Morbidelli, A.\ 2019.\ The role of dissipative evolution for three-planet, near-resonant extrasolar systems.\ \aap\ 625. doi:10.1051/0004-6361/201935259

\bibitem
{2013A&A...556A..28B} Batygin, K., Morbidelli, A.\ 2013.\ Analytical treatment of planetary resonances.\ \aap\ 556. doi:10.1051/0004-6361/201220907

\bibitem
{2018MNRAS.477.1414C} Charalambous, C., Mart{\'\i}, J.~G., Beaug{\'e}, C., Ramos, X.~S.\ 2018.\ Resonance capture and dynamics of three-planet systems.\ \mnras\ 477, 1414–1425. doi:10.1093/mnras/sty676

\bibitem
{2014A&A...566A.137D} Delisle, J.-B., Laskar, J., Correia, A.~C.~M.\ 2014.\ Resonance breaking due to dissipation in planar planetary systems.\ \aap\ 566. doi:10.1051/0004-6361/201423676

\bibitem
{2020MNRAS.494.4950P} Pichierri, G., Morbidelli, A.\ 2020.\ The onset of instability in resonant chains.\ \mnras\ 494, 4950–4968. doi:10.1093/mnras/staa1102

\bibitem
{2017A&A...601A..15L} Liu, B., Ormel, C.~W., Lin, D.~N.~C.\ 2017.\ Dynamical rearrangement of super-Earths during disk dispersal. I. Outline of the magnetospheric rebound model.\ \aap\ 601. doi:10.1051/0004-6361/201630017

\bibitem
{2017A&A...604A...1O} Ormel, C.~W., Liu, B., Schoonenberg, D.\ 2017.\ Formation of TRAPPIST-1 and other compact systems.\ \aap\ 604. doi:10.1051/0004-6361/201730826

\bibitem
{2008EAS....29..165M} Masset, F.~S.\ 2008.\ Planet Disk Interactions.\ EAS Publications Series 29, 165–244. doi:10.1051/eas:0829006

\bibitem
{2019A&A...627A.149S} Schoonenberg, D., Liu, B., Ormel, C.~W., Dorn, C.\ 2019.\ Pebble-driven planet formation for TRAPPIST-1 and other compact systems.\ \aap\ 627. doi:10.1051/0004-6361/201935607

\bibitem
{2022NatAs...6...72M} Morbidelli, A. et al.\ 2022.\ Contemporary formation of early Solar System planetesimals at two distinct radial locations.\ \nastro\ 6, 72–79. doi:10.1038/s41550-021-01517-7

\bibitem
{2014MNRAS.437...96F} Fendyke, S.~M., Nelson, R.~P.\ 2014.\ On the corotation torque for low-mass eccentric planets.\ \mnras\ 437, 96–107. doi:10.1093/mnras/stt1867

\bibitem
{2023A&A...670A.148P} Pichierri, G., Bitsch, B., Lega, E.\ 2023.\ A recipe for orbital eccentricity damping in the type-I regime for low-viscosity 2D discs.\ \aap\ 670. doi:10.1051/0004-6361/202245196 

\bibitem
{2021A&A...648A..69A} Ataiee, S., Kley, W.\ 2021.\ Pushing planets into an inner cavity by a resonant chain.\ \aap\ 648. doi:10.1051/0004-6361/202038772

\bibitem
{2006ApJ...642..478M} Masset, F.~S., Morbidelli, A., Crida, A., Ferreira, J.\ 2006.\ Disk Surface Density Transitions as Protoplanet Traps.\ \apj\ 642, 478–487. doi:10.1086/500967

\bibitem
{2013ApJ...778..169B} Batygin, K., Adams, F.~C.\ 2013.\ Magnetic and Gravitational Disk-Star Interactions: An Interdependence of PMS Stellar Rotation Rates and Spin-Orbit Misalignments.\ \apj\ 778. doi:10.1088/0004-637X/778/2/169

\bibitem
{2023ASPC..534..539M} Manara, C.~F. et al.\ 2023.\ Demographics of Young Stars and their Protoplanetary Disks: Lessons Learned on Disk Evolution and its Connection to Planet Formation.\ Protostars and Planets VII 534, 539. doi:10.48550/arXiv.2203.09930

\bibitem
{2017A&A...605A..96D} Delisle, J.-B.\ 2017.\ Analytical model of multi-planetary resonant chains and constraints on migration scenarios.\ \aap\ 605. doi:10.1051/0004-6361/201730857

\bibitem
{2018CeMDA.130...54P} Pichierri, G., Morbidelli, A., Crida, A.\ 2018.\ Capture into first-order resonances and long-term stability of pairs of equal-mass planets.\ \celmech\ 130. doi:10.1007/s10569-018-9848-2

\bibitem
{2021AJ....161..290S} Siegel, J.~C., Fabrycky, D.\ 2021.\ Resonant Chains of Exoplanets: Libration Centers for Three-body Angles.\ \aj\ 161. doi:10.3847/1538-3881/abf8a6

\bibitem
{2008MNRAS.387..747M} Michtchenko, T.~A., Beaug{\'e}, C., Ferraz-Mello, S.\ 2008.\ Dynamic portrait of the planetary 2/1 mean-motion resonance - I. Systems with a more massive outer planet.\ \mnras\ 387, 747–758. doi:10.1111/j.1365-2966.2008.13278.x

\bibitem
{2008A&A...482..677C} Cresswell, P., Nelson, R.~P.\ 2008.\ Three-dimensional simulations of multiple protoplanets embedded in a protostellar disc.\ \aap\ 482, 677–690. doi:10.1051/0004-6361:20079178

\bibitem
{2002ApJ...565.1257T} Tanaka, H., Takeuchi, T., Ward, W.~R.\ 2002.\ Three-Dimensional Interaction between a Planet and an Isothermal Gaseous Disk. I. Corotation and Lindblad Torques and Planet Migration.\ \apj\ 565, 1257–1274. doi:10.1086/324713

\bibitem
{2017ApJ...835..230F} Flock, M., Fromang, S., Turner, N.~J., Benisty, M.\ 2017.\ 3D Radiation Nonideal Magnetohydrodynamical Simulations of the Inner Rim in Protoplanetary Disks.\ \apj\ 835. doi:10.3847/1538-4357/835/2/230

\bibitem
{2017MNRAS.470.1750I} Izidoro, A. et al.\ 2017.\ Breaking the chains: hot super-Earth systems from migration and disruption of compact resonant chains.\ \mnras\ 470, 1750–1770. doi:10.1093/mnras/stx1232

\bibitem
{2021A&A...650A.152I} Izidoro, A. et al.\ 2021.\ Formation of planetary systems by pebble accretion and migration. Hot super-Earth systems from breaking compact resonant chains.\ \aap\ 650. doi:10.1051/0004-6361/201935336

\bibitem
{1966Icar....5..375G} Goldreich, P., Soter, S.\ 1966.\ Q in the Solar System.\ \icarus\ 5, 375–389. doi:10.1016/0019-1035(66)90051-0

\end{thebibliography}

\begin{thebibliography}{9}
\makeatletter
\addtocounter{\@listctr}{36}
\makeatother

\bibitem
{2019AJ....158..238H} Hadden, S.\ 2019.\ An Integrable Model for the Dynamics of Planetary Mean-motion Resonances.\ \aj\ 158. doi:10.3847/1538-3881/ab5287

\bibitem
{2023AJ....165...33D} Dai, F. et al.\ 2023.\ TOI-1136 is a Young, Coplanar, Aligned Planetary System in a Pristine Resonant Chain.\ \aj\ 165. doi:10.3847/1538-3881/aca327

\bibitem
{1979ApJ...233..857G} Goldreich, P., Tremaine, S.\ 1979.\ The excitation of density waves at the Lindblad and corotation resonances by an external potential..\ \apj\ 233, 857–871. doi:10.1086/157448

\bibitem
{1980ApJ...241..425G} Goldreich, P., Tremaine, S.\ 1980.\ Disk-satellite interactions..\ \apj\ 241, 425–441. doi:10.1086/158356

\bibitem
{1997Icar..126..261W} Ward, W.~R.\ 1997.\ Protoplanet Migration by Nebula Tides.\ \icarus\ 126, 261–281. doi:10.1006/icar.1996.5647

\bibitem
{2011MNRAS.418.1043Q} Quillen, A.~C.\ 2011.\ Three-body resonance overlap in closely spaced multiple-planet systems.\ \mnras\ 418, 1043–1054. doi:10.1111/j.1365-2966.2011.19555.x

\bibitem
{2020A&A...641A.176P} Petit, A.~C., Pichierri, G., Davies, M.~B., Johansen, A.\ 2020.\ The path to instability in compact multi-planetary systems.\ \aap\ 641. doi:10.1051/0004-6361/202038764

\bibitem
{1995CeMDA..62..193L} Laskar, J., Robutel, P.\ 1995.\ Stability of the Planetary Three-Body Problem. I. Expansion of the Planetary Hamiltonian.\ \celmech\ 62, 193–217. doi:10.1007/BF00692088

\bibitem
{2011MNRAS.410..293P} Paardekooper, S.-J., Baruteau, C., Kley, W.\ 2011.\ A torque formula for non-isothermal Type I planetary migration - II. Effects of diffusion.\ \mnras\ 410, 293–303. doi:10.1111/j.1365-2966.2010.17442.x

\bibitem
{2006Icar..181..587C} Crida, A., Morbidelli, A., Masset, F.\ 2006.\ On the width and shape of gaps in protoplanetary disks.\ \icarus\ 181, 587–604. doi:10.1016/j.icarus.2005.10.007

\bibitem
{2021ApJ...907...81L} Lin, Y.-C., Matsumoto, Y., Gu, P.-G.\ 2021.\ Formation of Multiple-planet Systems in Resonant Chains around M Dwarfs.\ \apj\ 907. doi:10.3847/1538-4357/abd0f3

\bibitem
{2022NatAs...6...80R} Raymond, S.~N. et al.\ 2022.\ An upper limit on late accretion and water delivery in the TRAPPIST-1 exoplanet system.\ \nastro\ 6, 80–88. doi:10.1038/s41550-021-01518-6

\bibitem
{2017MNRAS.464.3728B} Bolmont, E. et al.\ 2017.\ Water loss from terrestrial planets orbiting ultracool dwarfs: implications for the planets of TRAPPIST-1.\ \mnras\ 464, 3728–3741. doi:10.1093/mnras/stw2578

\bibitem
{2020ApJ...889...77H} Hori, Y., Ogihara, M.\ 2020.\ Do the TRAPPIST-1 Planets Have Hydrogen-rich Atmospheres?.\ \apj\ 889. doi:10.3847/1538-4357/ab6168

\bibitem
{2023NatAs...7..330B} Batygin, K., Morbidelli, A.\ 2023.\ Formation of rocky super-earths from a narrow ring of planetesimals.\ \nastro\ 7, 330–338. doi:10.1038/s41550-022-01850-5

\bibitem
{2023Natur.618...39G} Greene, T.~P., Bell, T.~J., Ducrot, E., Dyrek, A., Lagage, P.-O., Fortney, J.~J.\ 2023.\ Thermal emission from the Earth-sized exoplanet TRAPPIST-1 b using JWST.\ \nat\ 618, 39–42. doi:10.1038/s41586-023-05951-7

\bibitem
{2023Natur.620..746Z} Zieba, S. et al.\ 2023.\ No thick carbon dioxide atmosphere on the rocky exoplanet TRAPPIST-1 c.\ \nat\ 620, 746–749. doi:10.1038/s41586-023-06232-z

\bibitem
{2021A&A...649L...5B} Bitsch, B., Raymond, S.~N., Buchhave, L.~A., Bello-Arufe, A., Rathcke, A.~D., Schneider, A.~D.\ 2021.\ Dry or water world? How the water contents of inner sub-Neptunes constrain giant planet formation and the location of the water ice line.\ \aap\ 649. doi:10.1051/0004-6361/202140793

\bibitem
{2006ApJ...648..696I} Ikoma, M., Genda, H.\ 2006.\ Constraints on the Mass of a Habitable Planet with Water of Nebular Origin.\ \apj\ 648, 696–706. doi:10.1086/505780

\bibitem
{2012A&A...546A..71D} Delisle, J.-B., Laskar, J., Correia, A.~C.~M., Bou{\'e}, G.\ 2012.\ Dissipation in planar resonant planetary systems.\ \aap\ 546. doi:10.1051/0004-6361/201220001

\bibitem
{2022A&A...658A.184O} Ogihara, M., Kokubo, E., Nakano, R., Suzuki, T.~K.\ 2022.\ Rapid-then-slow migration reproduces mass distribution of TRAPPIST-1 system.\ \aap\ 658. doi:10.1051/0004-6361/202142354


\end{thebibliography}
\end{document}